\documentclass[11pt,oneside,reqno]{amsart}

\usepackage{amsmath}
\usepackage{amsthm,amsfonts}
\usepackage{amssymb,latexsym}

\usepackage{color}
\usepackage{multirow}
\usepackage[boxed]{algorithm2e}
\usepackage{subfig}

\usepackage{bm}

\usepackage{upref}
\usepackage[utopia]{mathdesign} 
\usepackage{hyperref}
\usepackage{textcase}

\usepackage{rotating}

\allowdisplaybreaks[4]  

\usepackage[letterpaper,left=1.40in,right=1.40in,top=1.25in,bottom=1.25in]{geometry}

\begin{document}

\title{On wavelet to select the parametric form of a regression model}

\author{Eufr\'{a}sio de Andrade Lima Neto}
\address{Departamento de Estat\'{\i}stica\\
         Universidade Federal da Para\'{\i}ba\\
     Cidade Universit\'{a}ria\\
         Jo\~{a}o Pessoa/PB, 58059--900, Brazil}
\email[E.A.\ Lima Neto]{eufrasio@de.ufpb.br}
      
\author{Alu\'{i}sio de Souza Pinheiro}
\address{Departamento de Estat\'{\i}stica\\
         Universidade Estadual de Campinas\\
     IMECC\\
         Campinas/SP, 13083--859, Brazil}

\author{Adenice Gomes de Oliveira Ferreira}
\address{Departamento de Estat\'{\i}stica\\
	Universidade Federal da Para\'{\i}ba\\
	Cidade Universit\'{a}ria\\
	Jo\~{a}o Pessoa/PB, 58059--900, Brazil}


\date{\today}


\begin{abstract}
	\textcolor{black}{Let $Y$ be a response variable related with a set of explanatory variables and let $f_1, f_2, \ldots, f_k$ a set of the parametric forms representing a set of candidate's model. Let $f^{*}$ be the true model among the set of $k$ plausible models. We discuss in this paper the use of wavelet regression method as auxiliary for the choice of the ``true'' parametric form of a regression model, particularly, for the cases of nonlinear regression and generalized linear models. The use of a non-parametric method for the choice of the more appropriate parametric equation in regression problems would be interesting in practice due to the simplicity and because the probabilistic assumptions are not required.
	We evaluate the performance of the proposed wavelet procedure based on the true classification rate of the correct parametric form among a range of $k$ candidate models, taking into account a wide ranges of scenarios and configurations as well as in real data set applications.
	}
\end{abstract}

\maketitle

\section{\textcolor{black}{Introduction}}

 Parametric regression models are widely used in many fields and represent one of the most important statistical tools. The generalized linear models (GLMs) represent one of the most important developments in statistical theory over the past several decades\cite{NW1972}. A GLM is characterized by three terms. The first is the random component with the response variable belonging to the exponential family of distributions. The second is the systematic component represented by a linear predictor that includes the explanatory variables. The third term is the link function which connects the linear predictor to the response variable mean. Another important topic in statistical modeling is the nonlinear regression, with a large applicability in several fields like biology, engineering, medicine, among others \cite{BW2007,RS2008}.

Wavelets have been developed in functional analysis as bases for $L_2(\bm{R})$, as well as some of its subspaces. These classes of functions contain a large number of diverse elements, which makes them suitable for broad theoretical and numerical applications. For instance, they form unconditional bases for some large functional classes, which leads to optimal estimators and tests \cite{Vidakovic1999}. 

An important step in the employment of some parametric regression model is the choice the mathematical function or the regression equation that relates the response variable $Y$ with a set of explanatory variables $X_1, \ldots, X_p$. 
In the framework of the GLM this step represents the choice of the link function. This function defines the regression equation that relates the random component to the linear predictor. Link misspecification can lead to several problems on a GLM application, such as  bias in the regression parameters and in the mean response estimates \cite{CS1992,CR2006}. A methodology which finds an appropriate link function for a GLM is still an open problem. Techniques have been proposed to evaluate if a predetermined link function is adequate for a fitted GLM  \cite{B1996, CM1989, H1985, P1980}. A scatter plot between the fitted response variable ($\hat{y}$) and the fitted linear predictor  ($\hat{\eta}$) represents an informal procedure to verify whether the link function is suitable. Thus, most of the current techniques are straightforward adaptations from linear models' procedures.

The same problem occurs to define the ``best'' nonlinear function in the framework of nonlinear regression. Usually, it is recommended the previously knowledge of the nonlinear relationship between the response and the explanatory variables. However, in practice, this is not always possible and the researcher not have information about the true nonlinear model. Exploratory techniques are used to detect the more appropriate nonlinear function among a range of eligible nonlinear functions.

Wavelet methods have been used within parametric models in several instances. The references \cite{Carl2008,Carl2010} study the employment of wavelet methods to remove the effects of spatial auto-correlation in generalized linear models while 
\cite{Fadili2004} applies penalized partially linear models to fMRI data, and \cite{Gannaz2007} discusses the wavelet application in partially linear models aiming robust estimation. Wavelet techniques have been successfully employed in the analysis of linear normal regression models under long range dependence by \cite{Fadili2002}. 

\textcolor{black}{In this paper, we propose to consider a wavelet regression (WR) model as alternative way to find the best parametric equation for nonlinear regression and generalized linear model problems. The aim is to verify the accuracy of the WR to identity the true nonlinear function or the true link function in a wide range of scenarios}. 

\textcolor{black}{The paper is organized as follows: section 2 presents an overview about Wavelets and the section 3 brings a brief description about the parametric regression methods GLM and nonlinear regression. Section 4 exhibits the Monte Carlo experiments and evaluates the performance of the wavelet procedure to detect the true parametric form based on a wide rage of scenarios. Section 5 brings applications to real data set. Finally, section 6 closes the text with some concluding remarks. The {\tt R} code is available in the supplementary material.}

\section{Wavelets} \label{WaveSec}

The theory of wavelets can be traced back to the beginning of the 1900's but the approach which unifies all the varying concepts behind this theory as a viable tool for data analysis is the so-called Multi-Resolution Analysis \cite{Mallat1989}. 
We direct the readers to \cite{Daubechies1992}, \cite{Vidakovic1999} and \cite{Morettin2016} for a thorough review of wavelets, from the mathematical and statistical points of view. 

A Multi-Resolution Analysis (MRA) in $L_2(R)$ is a nested sequence of closed subspaces, $\{V_j\}_{j\in Z}$ with four basic properties:
\begin{list}{}{}
	\item i - {\bf Hierarchy}
	\begin{equation}\label{C4eq31} \nonumber
	V_j \subset V_{j+1} \subset L_2{(R)} ~~ \forall j \in
	{Z}
	\end{equation}
	\item ii-  {\bf Dense Union and Trivial Intersection}
	\begin{equation} \label{C4eq32} \nonumber
	\overline{\bigcup_{j \in {Z}}V_j}=L_2{(R)}  and
	\bigcap_{j \in {Z}}V_j= \{0\}
	\end{equation}
	\item iii- {\bf Self-Similarity}
	\begin{equation} \label{C4eq33} \nonumber
	m(2^jt) \in V_j \Leftrightarrow m(t) \in V_0  ~\forall j \in
	{Z}
	\end{equation}
	\item iv - {\bf Natural Basis}
	$\exists\phi$ $\in$ $V_0$ so that $T^k \phi(t)$
	$=\phi(t-k)$ $\forall k$ $\in$ ${Z}$ spans $V_0$, i.e., 
	\begin{equation}\label{C4eq34}
	V_0= \left \{ m \in L_2{(R)} ~|~ f(t) = \sum_{k \in
		{Z}} c_k \phi(t-k) \right \}
	\end{equation}
	for some appropriate sequence  $\{c_k\}_{k\in Z}$. $\{ \phi( \cdot
	- k), k \in {Z} \}$ is called an orthonormal basis of $V_0$.
\end{list}

\noindent $\phi(\cdot)$ is called a scale function or father-wavelet. It generates other bases by translation and dilation: $\phi_j(t) = 2^{j/2} \phi(2^jt-k)~~j
\in {Z}~~k \in {Z}$. The orthogonal system $\phi_{j,k}(\cdot)$ spans $V_j$ for each $j$, i.e.,
\begin{equation}\label{C4eq36}
V_j= \left \{ m \in L_2{(R)} ~|~ f(t) = \sum_{k \in{Z}} \alpha_{j,k} \phi_{j,k}(t) \right \},   \forall
j \in{Z}
\end{equation}
for some sequence $\{\alpha_{j,k}\}_{k \in {Z}}$, where 
$\{ \phi_{j,k}(\cdot), k \in{Z} \}$  is an orthonormal basis for  $V_j$ and $\alpha_{j,k}=<m,\phi_{j,k}>_{L_2}$. Any $m(\cdot)$ in $L_2(R)$ can be written as
\begin{equation}\label{C4eq37} \nonumber
m(t)=\lim_{j\rightarrow\infty} \sum_{k \in
	{Z}}\alpha_{j,k}\phi_{j,k}(t)
=\lim_{j\rightarrow\infty}P_j m(t),
\end{equation}
where  $P_j m(t)$ is the orthogonal projection of   $m$ on $V_j$ . It is easy to see that
$\lim_{j\rightarrow -\infty}P_j m(t)=0$ and  $\langle \phi_{j,b},\phi_{j,a} \rangle_{L_2} =
\int_{-\infty}^{+\infty} \phi_{j,b}(t)\overline{\phi_{j,a}(t)}dt =
\delta_b^a$, where $\delta_b^a = 0 $ if $a \neq b$, and  $\delta_b^a = 1 $ if $a =
b$.  The reason for the broad applicability of wavelets is given by the associated filters with nice numerical properties such that:
\begin{equation}\label{C4eq40} \nonumber
\phi(t) = \sum_{k \in {Z}}h_k \phi_{1,k}(t) = \sum_{k \in
	{Z}}h_k \sqrt{2} \phi(2t-k),
\end{equation}
where  $h_k= \sqrt{2} \int_{R} \phi(t)
\phi(2t-k) dt , k \in {Z}$ is known as a scale function filter.

A Multi-resolution Analysis (MRA) of $L_2(R)$ is called $r$-regular, $r$ $\in{N}$, if the scale function $\phi(\cdot)$, defined by (\ref{C4eq34}), is such that:
\begin{equation}\label{C5regul} \nonumber
|\phi^{(k)}(t)| \leq \frac{C_m}{(1 + | t | )^{m}},  ~~\forall
k \leq r ~~ \forall k \in {N} ~~ \forall m \in
{N}.
\end{equation}

Another filter $g_k$  is defined from $h_k$ via the  so-called mirrored quadrature relation (QMF): $g_n =(-1)^{n}h_{1-n}$. We can write $g_k= \sqrt{2} \int_{{R}} \psi(t) \phi(2t-k) dt ~~\forall k \in {Z}$ and $\{\psi_{j,k}(t)=2^{\frac{j}{2}}\psi(2^jt-k),  j\in{Z}, k \in
{Z}\}$ spans $L_2(R)$ as well. Let $W_j= \left \{ m \in L_2{(R)} /~|~m(t) \stackrel{L_2} =
\sum_{k \in {Z}} \beta_{j,k} \psi_{j,k}(t) \right \}$. Then,  
$V_{j+1}=V_j \oplus W_j,   \forall j \in{Z}$ and
\begin{equation}\label{C4eq52} \nonumber
L_2{(R)}= \overline{\bigoplus_{j \in {Z}} W_j}.
\end{equation}

\noindent Thence, any function $m\in L_2{(R)}$ can be written in $L_2$-sense as:
\begin{equation}\label{C4eq53}
m(t) = \sum_{j \in{Z}}\sum_{k \in
	{Z}}\beta_{j,k}\psi_{j,k}(t) \nonumber = \sum_{k \in
	{Z}}\alpha_{j_0,k}\phi_{j_0,k}(t) + \sum_{j \geq j_0 \in
	{Z}}\sum_{k \in {Z}}\beta_{j,k}\psi_{j,k}(t),
\end{equation}
for an arbitrary $j_0$. The choice of the wavelet basis depends on several aspects. The wavelets regularity is very important for statistical optimality, and can be assessed by the number of null moments:
\begin{equation}\label{C4eq64}
\bm{M}_k = \int_{{R}} t^k \psi(t) dt.
\end{equation}

\noindent But $\phi$ and $\psi$ have $N$ null moments if and only if 
\begin{equation}\label{C4eq68} \nonumber
\sum_{n \in {Z}} n^k g_n = \sum_{n \in {Z}} n^k
(-1)^n h_n = 0,  \mbox{ for }k=0,1,...,N-1.
\end{equation}

In general, filters have an infinite number of non-null terms. Two special classes are given by: {\it N-regular} MRA's, i.e. with $N$ null moments; and by compactly supported wavelets. In both cases, the number of non-null terms is $2N$ \cite{Daubechies1992}. One such family of compactly supported wavelets is the Daubechies family of wavelets, and a particular case  is the Haar basis, also considered the {\it first} wavelet, defined by $\phi(t) = 1_{[0,1]} (t)$ and $ \psi(t) = 1_{[0,{1}/{2}]} (t)- 1_{[{1}/{2},1]} (t)$, or, by its filtration $ h_0 = h_1 = {\sqrt{2}}/2$ and $g_0 ={\sqrt{2}}/2$, $g_1
=-{\sqrt{2}}/2$. 

The Daubechies' are indexed by the number of null moments $N$ as 
{\it Daubechies(N)}, with support $[0,2N-1]$ and associated filters of length $2N$. For instance, one has for the {\it Daubechies(2)}, 
\begin{equation} \nonumber
h_0 = \frac{1 + \sqrt{3}}{4 \sqrt{2}},  h_1 = \frac{3 +
	\sqrt{3}}{4 \sqrt{2}},  h_2 = \frac{3 - \sqrt{3}}{4
	\sqrt{2}},  h_3 = \frac{1 - \sqrt{3}}{4 \sqrt{2}}.
\end{equation}

The Daubechies wavelets do not have, other than in the Haar case, closed forms. For this reason, we employ the  Daubechies-Lagaria {\it Cascade} Algorithm, which  allows the computation of any $\phi(t)$ for $t \in {R}$, with any predetermined precision. Consider $\phi(\cdot)$ the scale function for the {\it Daubechies(N)} basis and $\{h_k\}_{k \in {R}}$ its associated filter. For any $t \in (0,1)$ and $\{ d_1,d_2,...\}$ the dyadic representation of $t$, defined by  $t=
\sum_{j=1}^{\infty} d_j 2^{-j}$, we define the matrices $T_0$ and $T_1$ as:
\begin{equation}\label{C4eq79}
T_0 = ( \sqrt{2} h_{2i-j-1})_{1 \leq i,j \leq 2N-1}  T_1 = (
\sqrt{2} h_{2i-j})_{1 \leq i,j \leq 2N-1}.
\end{equation}
Then, $\lim_{n \rightarrow \infty} T_{d_{1}}...T_{d_{n}} $
\begin{equation}\label{C4eq80}
= \left [
\begin{array}{cccc}
\phi(t) & \phi(t) & \cdots & \phi(t) \\  \phi(t+1) & \phi(t+1) & \cdots
& \phi(t+1) \\ \vdots & \vdots& \ddots & \vdots \\ \phi(t+2N -2) & \phi(t+2N -2)
& \ldots & \phi(t+2N -2) \\ .
\end{array} \right ].
\end{equation}

The class of square integrable functions is in general too large and diverse to be of interest in practice. But there are smaller spaces which are large enough to be useful in a good number of problems but still possess regularity conditions which are relevant. Two such subspaces are H\"older and Besov spaces, say $\mathbf{H}_{\alpha}(R)$ and $\mathcal{B}_{p,q}^s$.

One has that some function $m$ belongs to  $\mathcal{H}_{\alpha}(R)$  (or $\mathcal{B}_{p,q}^s$) if and only if its wavelet coefficients follow a certain decay law. The wavelet basis is then called an unconditional basis for  $\mathcal{H}_{\alpha}(R)$  (or $\mathcal{B}_{p,q}^s$). In applications this property results in the analysis of the estimated coefficients in order to assess the degree of regularity the data possess. This leads to empirical coefficients shrinkage and to the optimality of wavelet-based estimation and test procedures in minimax sense \cite{Vidakovic1999,Morettin2016}.

Wavelets can be used as building blocks of $L_2(R^p)$ (or suitable multidimensional functional sub-classes). There are several constructions, each being more, or less,  interesting depending on the researcher's goals \cite{Morettin2016}. We use here the most direct and mathematically more appealing MRA. Its basis is taken as the tensor product of all the one-dimensional bases. For instance, for $L_2(R^2)$, we have the one-dimensional MRA approximation and wavelet spaces of scale $j$ given by $V_j$ and $W_j$. Its bases are given by $\{\phi_{j,k},~~k\in Z\}$ and $\{\psi_{j,k},~~k\in Z\}$, respectively. The MRA for $L_2(R^2)$ is such that its approximation and wavelet spaces are given by:
$V_j=span\{\phi^{(2)}_{jk}, ~~k\in Z\}$ and $W_j=span\{\psi^{(1)}_{jk}(x,w),\psi^{(2)}_{jk}(x,w), \psi^{(3)}_{jk}(x,w),~~k\in Z \}$, where $\phi^{(2)}_{jk}(x,w)=\phi_{jk}(x)\phi_{jk}(w)$, $\psi^{(1)}_{jk}(x,w)=\phi_{jk}(x)\psi_{jk}(w)$, $\psi^{(2)}_{jk}(x,w)=\psi_{jk}(x)\phi_{jk}(w)$ and $\psi^{(3)}_{jk}(x,w)=\psi_{jk}(x)\psi_{jk}(w)$. We should note that each wavelet has a different purpose, in the sense that $\psi^{(1)}$, $\psi^{(2)}$ and $\psi^{(3)}$ capture changes in horizontal, vertical or diagonal fashion, respectively. The extension to higher dimensions is straightforward.

We employ as WM the wavelet regression estimator proposed by \cite{Kovac2000}. The idea is to apply wavelet regression to non-equally spaced data sets. First the grid points are defined as $\tilde{t}_k=(K+1/2)2^{-J}$, where $k\in\{0,\ldots,2^J-1\}$. The gridded response values are then calculated as $\tilde{y}_k$ by a linear transformation of the original $y$'s. We simply use as $\tilde{y}_k$ the observation(s) which lies on $[k2^{-J},(k+1)2^{-J}]$. Whenever no observation can be found on a grid interval, we take the nearest observation to the left of it. In this way we transform a non-equally spaced data to an equally spaced data and,  moreover, this is done in such a way as to produce a sample size which is a power of $2$. Hence, usual DWT techniques can be employed. Thresholding is performed on the estimated coefficients and we write the WM estimator as
\begin{equation} \label{WM}
\hat{m}(x)=\sum_{k=0}^{2^{j_0}-1}\hat{c}_{j_0k}\psi_{j_0k}(x)+\sum_{j\geq j_0}^{J_max-1}\sum_{k=0}^{2^j}\hat{d}_{jk}^{thr}\psi_{jk}(x),
\end{equation}
where $\hat{c}_{j_0k}$ are the estimated approximation coefficients and $\hat{d}_{jk}^{thr}$ are the thresholded detail coefficients for the $j$-th scale \cite{Kovac2000}.

\section{\textcolor{black}{Parametric regression models background}}

A parametric regression model involves a dependent variable $Y$, a set of explanatory variables $X_1, X_2, \ldots, X_p$ and a vector of unknown parameters $\bm{\beta}$ that need to be estimated. The relationship between $Y$ and  $X_1, X_2, \ldots, X_p$ is given through a function $f$ that must be specified. Thus, a regression model that relates the response and explanatory variables is defined by
\begin{equation}
Y = f(\mathbf X, \bm{\beta}) + \epsilon,
\end{equation}
\noindent where $\epsilon$ is the error of the model that follows a probability distribution. The form of the function $f$ is based on knowledge about the relationship between $Y$ and $\mathbf X$ that does not rely on the data. However, if no such knowledge is available, a flexible or convenient form for $f$ need to be specified.
The choice of the parametric form $f$ represents an important step in the model choice, particularly, in the class of generalized linear models and in the nonlinear regression models.

\subsection{Generalized linear model} 

Let $Y = \{y_1,\ldots,y_n\}$ be a set of observations that represents a random sample of the response  variable $Y$. We consider that the density probability function of $Y$ belongs to the exponential family of 
distributions if its probability mass function has the following form:
\begin{equation}\label{random}
f( y,\theta,\phi)=
exp\left[a(\phi)^{-1}\{y\theta  - b(\theta)\}+ c(y,\phi)\right].
\end{equation}
The functions $a(\cdot), b(\cdot)$ and $c(\cdot)$ are known, $\theta$ is the canonical parameter and $\phi$ is a nuisance parameter. The mean and variance of $Y$  can be obtained from well-known equations of natural exponential families. The log-likelihood function for the $i$th observation can be written as
\begin{equation}\label{loglikelihood}
l_{i}=l_{i}(\theta,\phi,y_i) =
a(\phi)^{-1} \{y_{i}\theta -b(\theta)\}+c(y_{i},\phi).
\end{equation}

A parametric regression model based on the GLM framework consists of two parts: a random and a syste\-ma\-tic component. The former considers the response variable $Y$ having a distribution from the exponential family (\ref{random}). 
In the systematic component, the explanatory variables $X_{1},\ldots,X_p$ 
are responsible for the variability of $Y$, being defined by
\begin{eqnarray}\label{systematic}
\bm{\eta} =
g(\bm{\mu})=  \mathbf
X\bm{\beta},
\end{eqnarray}
where $\mathbf X$ is the design matrix
formed by the observed values of the explanatory variables $X_{1}, X_{2}, \ldots, X_{p}$,  $\bm{\beta}$ is the vector of parameters,
$\bm{\eta}$ is the vector of linear predictors, $\bm{\mu}$ is the vector of means of $Y$, i.e., with
$\bm{\eta}$=$(\eta_{1},\ldots,
\eta_{n})^{T},\bm{\mu}$ = $(\mu_{1},\ldots,
\mu_{n})^{T}$ and $\bm{\beta}$=$(\beta_{0},\ldots,\beta_{p})^T$. The link function, call it $g(\bm{\mu})$, connects the res\-pon\-se variable mean to the explanatory variables. If $Y$ is continuous, a few functions available for a \textbf{GLM} are: the \textit{identity, logarithmic, inverse, power}.
Some link functions have nice properties and may be preferred
in some particular situations. These are called the {\it canonical} link functions and occur when the canonical parameter equals the linear predictor, i.e. if $\bm{\eta}=\bm{\theta}$

The maximum likelihood (ML) method is used as the theoretical basis for the estimation of $\bm{\beta}$, without the knowledge of $\phi$. Although $\phi$ can also be estimated by maximum likelihood there may be practical
difficulties for some exponential family distributions. A simple way to estimate $\phi$ is based on the deviance of the model. More about the GLM's can be found in \cite{NW1972, McCullagh1989}.

\subsection{\textcolor{black}{Nonlinear regression}}

The nonlinear regression model supposes that relation between the dependent and independent variable(s) occurs through a function that is a nonlinear combination of model parameters and depends on one or more independent variables.

The simple nonlinear regression is defined by
\begin{eqnarray}
y_i=f(x_i,\bm{\beta}) +\varepsilon_i,\quad i=1,2,...,n, \label{nonlinear}
\end{eqnarray}
\noindent where $y_i$ represents $i$-th value of the response variable $Y$, $f$ is a nonlinear and differentiable function related to the model parameters, $x_i$ is the $i$-th value of the independent variable  $X$, $\bm{\beta}$ is the vector of unknown parameters and $\varepsilon_i$, o $i$-th value of the unobserved error.

We assume that the error are random variables i.i.d following a normal distribution with mean $\mu_\varepsilon=0$ and variance  $\sigma_\varepsilon^2$. According with the equation (\ref{nonlinear}) is possible to claim that the simple linear regression model is a particular case of the simple nonlinear regression model, where the function $f(x_i,\beta)$ is the identity and $y_i$ is, consequently, given by $y_i=\beta_0 + \beta_1 x_i + \varepsilon$.
The nonlinear regression models are applied in several field like ecology, agriculture, biology, among others. For example, a widely used nonlinear function in biochemistry, in the study of enzymatic kinetics, is the nonlinear equation proposed by \cite{MM1913}. 
%

The normal equations are nonlinear being necessary the use of an iterative procedure to obtain the solution of the equation system \cite{BW2007}. 
Thus, to start the iterative process, is required to consider started values for the vector of parameters $\bm{\beta}$. The procedure is finished based on the convergence of the objective function or when the maximum number of iterations is reached. In the nonlinear regression model, the objective function is represented by the sum of squares of error, given by:
\begin{eqnarray}
SQE_{MNL}(\bm{\beta})=\sum_{i=1}^{n}(y_i-f(x_i,\bm{\beta}))^2. \label{eq3}
\end{eqnarray}

\noindent Some optimization methods can be used to obtain the parameter estimates that minimizes (\ref{eq3}), as for example, Gauss-Newton method, Conjugated gradient, LevenbergMarquardt method or BFGS (Broyden-Fletcher-Goldfarb-Shanno's Algorithm). Details about these method can be found in \cite{ChongZak2013}. 


Another important point related to nonlinear regression problems is the choice of the ``best'' nonlinear function $f$, that in many situations is not known. The Akaike information criterion (AIC) or a cross-validation procedure  are alternative to find the more appropriate function $f$ among a set of candidate functions \cite{CB2013, shao1993}. However, this problem continues open and the aim of the paper is to propose a new procedure to identify the best parametric model based on a nonparametric technique.

\section{\textcolor{black}{Wavelet procedure to identify the best parametric equation}}

Let $Y$ be a response variable related with almost one of a set of explanatory variables $X_1, X_2, \ldots, X_p$.
Let $f_1, f_2, \ldots, f_k$ be a set of candidate nonlinear functions or let $g_1, g_2, \ldots, g_k$ be a set on candidate link functions in a GLM context. Let $f^{*}$ (or $g^{*}$) be the true function among the set of $k$ plausible or candidate models. The aim is to use a wavelet regression (WR) model to find the best nonlinear function $f^{*}$  or the best link function $g^{*}$ that relates $Y$ with $X_1, X_2, \ldots, X_p$.

Our proposal is to compare the predicted values provided by the WR with each one of the predicted values provided by a set of $k$ eligible models $M_1, M_2, \ldots, M_k$ considering a performance error measure like, for example, root mean square error ($RMSE$) or median absolute error ($MAE$). The model with the lower value of $RMSE$ and/or $MAE$ will be considered the more appropriated parametric equation among all $k$ the candidate models.

The algorithm below describes a procedure that find the best (or true) parametric model comparing the fitted values of all $k$ candidate models and identifying the model more close of the fitted values provided by the WM.

\begin{algorithm}[H] \label{algo}
	\SetKwData{Left}{left}\SetKwData{This}{this}\SetKwData{Up}{up}
	\SetKwFunction{Union}{Union}\SetKwFunction{FindCompress}{FindCompress}
	\SetKwInOut{Input}{input}\SetKwInOut{Output}{output}
	\Input{$\mathbf{X}$, $\mathbf{y}$ and a set of $M_m$ parametric models, $m=1, 2, \ldots, k$.}
	\Output{A model $M^{*}$ with the $\min(RMSE_m)$}
	\BlankLine
	\emph{Initialization:}\newline
	\qquad Set $\bm{\tilde{\beta}} = (\mathbf X^\top \mathbf X )^{-1}\mathbf X^\top \mathbf y$\tcp*[r]{a start value for $\bm{\beta}$}
	\qquad Compute $\bm{\tilde{\eta}} = \mathbf X\bm{\tilde{\beta}}$\;
	\qquad Compute $\bm{\tilde{\eta}}_{*} = (\bm{\tilde{\eta}}- \min{(\bm{\tilde{\eta}})}/(\max{(\bm{\tilde{\eta}})} - \min{(\bm{\tilde{\eta}})})$ \tcp*[r]{re-scaled $\bm{\tilde{\eta}}$} 
	\qquad Compute $\tilde{\bm{\mu}} = m(\bm{\tilde{\eta}}_{*})$\tcp*[r]{fit a wavelet regression as in Eq.~\eqref{WM}} 
	\qquad Store $\tilde{\bm{\mu}}$
	\vskip2mm
	\emph{Fitting steps:}\newline	
	\For{$m = 1, \ldots, k$}{
		\hskip2mm Compute $\bm{\hat{\mu}_{m}} = f_{m}(\mathbf X; \bm{\hat{\beta}})$\tcp*[r]{fitted values of the candidate model m} 
		\hskip2mm Compute $RMSE_m = \sqrt{\frac{\sum_{i=1}^{n}(\mu_{i}^{m}-\tilde{\mu}_i)^{2}}{n}}$\tcp*[r]{performance error measure of the candidate model m} 
	}
	\qquad Return a Model $M^{*}$ with the $\min(RMSE)$
\end{algorithm}\DecMargin{1em}
\vskip0.5cm

The algorithm starts from an initial parametric solution (OLS) for the parameters vector $\bm{\beta}$. Notice that $\bm{\tilde{\eta}}$ is used to 
to build a non-equidistant grid based on the transformed linear predictor $\bm{\tilde{\eta}}_{*}$. The wavelet regression to $Y$ will be fit taking into account this non-equidistant grid $\bm{\tilde{\eta}}_{*}$. Another important aspect is that $\bm{\tilde{\eta}}_{*}$ allows to consider a wavelet model over $Y$ even when the number of explanatory variables $p > 1$. Finally, the fitted values of the WM are stored in $\bm{\tilde{\mu}}$.

In the fitting step we consider all the $k$ candidate models and compute the fitted values for each model. Thus, we consider the root mean square error (or another performance measure) as criterion to select the parametric model more close to the wavelet regression. This model, named as $M^{*}$, will be considered the best parametric approach to relate $Y$ with the set of explanatory variables. 

\section{Experimental evaluation} \label{EA}

\textcolor{black}{This section presents a Monte Carlo simulation study to evaluate the proposed wavelet procedure (WP) to identify the true parametric form of a regression model in the context of GLM and nonlinear regression.} 

\textcolor{black}{We will assess the true classification rate of the parametric form of a nonlinear function $f$ and the true classification rate of the link function $g$ in a GLM model with continuous distribution, take into account a wide range of scenarios.}

\textcolor{black}{The first scenario evaluate the proposed procedure for each one of four different true nonlinear functions, taking into account tree different dependence levels and sample sizes. In the second scenario we consider quite similar nonlinear functions. The aim is to evaluate the WP when there is almost one nonlinear function quite similar to the true nonlinear regression equation. We also consider as performance measure the true classification rate. Scenario 3 is similar to the scenario 1, but in the context of GLM. Finally, scenario 4 compares the predictive performance of the wavelet regression model against the true fitted GLM model. In this scenario, the aim is to verify if the nonparametric approach presents better fitted values when compared with the fitted values of the true fitted parametric model.  Below, we give details about the four simulation scenarios.}

\subsection*{\textcolor{black}{Scenario 1: identifying the more appropriate parametric form for a nonlinear regression model}}
\textcolor{black}{The scenario 1 evaluates the performance of the WP to identify the true nonlinear function considering synthetic data sets. The artificial data sets consider a predefined (true) nonlinear relationship between the response variable $Y$, the model parameters and the explanatory variable \textbf{X}. We select four different true nonlinear regression equation that are described below:
	\begin{eqnarray}
	y=f_1(x,\beta)=\frac{\beta_1}{\beta_2+e^{\beta_3x}} + \epsilon; \label{medicine}
	\end{eqnarray}
	\begin{eqnarray}
	y=f_2(x,\beta)=\beta_1+e^{-\beta_2x} + \epsilon; \label{industrial}
	\end{eqnarray}
	\begin{eqnarray}
	y=f_3(x,\beta)=\frac{\beta_2x}{\beta_1+x} + \epsilon; \label{chemistry}
	\end{eqnarray}
	\begin{eqnarray}
	y=f_4(x,\beta)=\beta_1\cos(2x)+\beta_2\sin(x) + \epsilon. \label{sincos}
	\end{eqnarray}
	\noindent The expression (\ref{medicine}) represents the logistic function with a large applicability in practical problems related to Medicine and Healthy. The expression (\ref{industrial}) represents an exponential function, applicable to industrial problems. The relationship (\ref{chemistry}) is well known in the Chemistry field and the function $f_4$ differs of the others due to the trigonometric arguments ``$\sin$'' and ``$\cos$''.}

\begin{figure}[htb]
	\centering
	\caption{Scenario 1 - true nonlinear regression models $f_1$-$f_4$ for sample size $n=128$ and strong dependence degree between $Y$ and $X$}\label{figuras}
	\includegraphics[width=5cm]{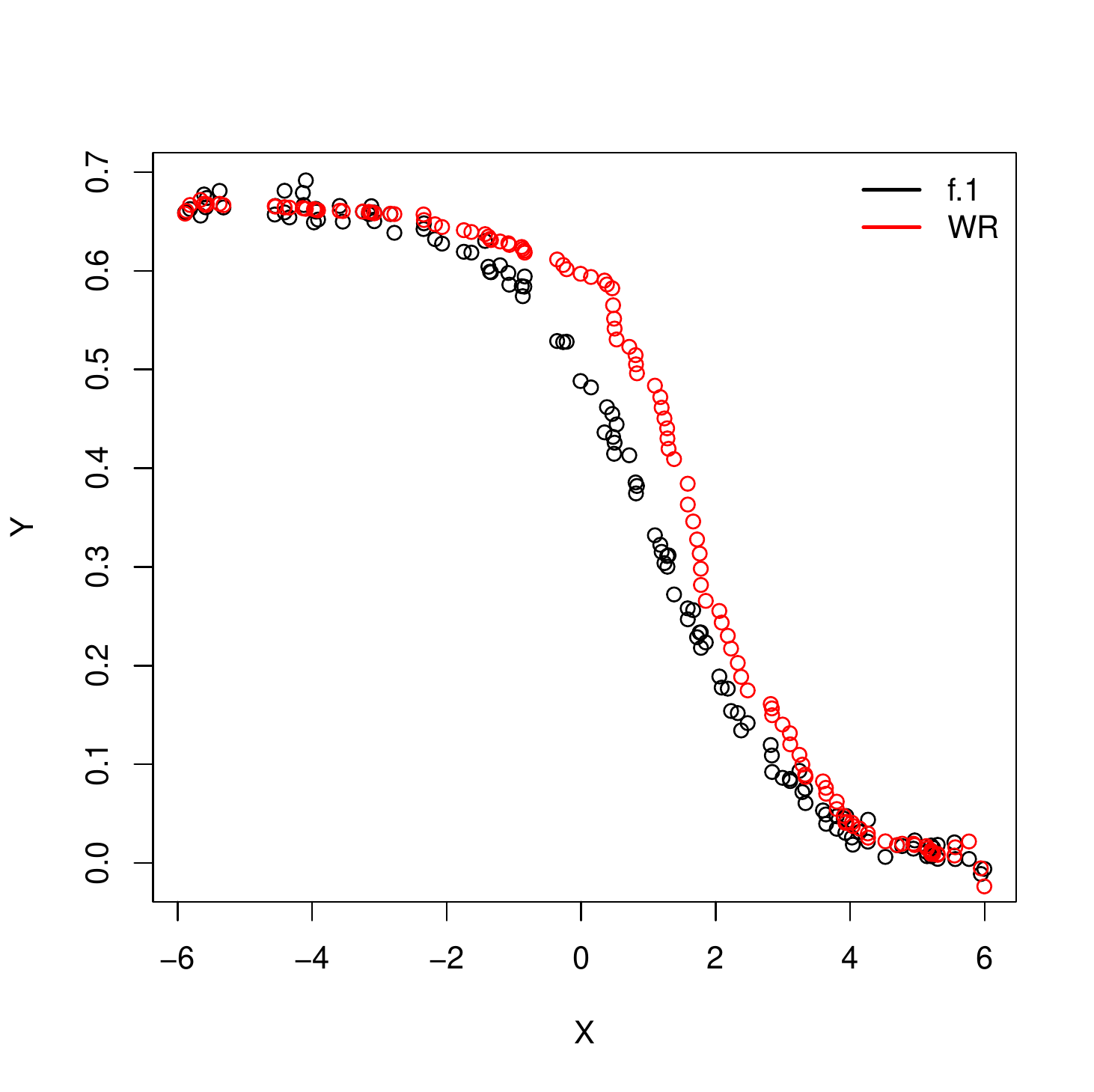}\includegraphics[width=5cm]{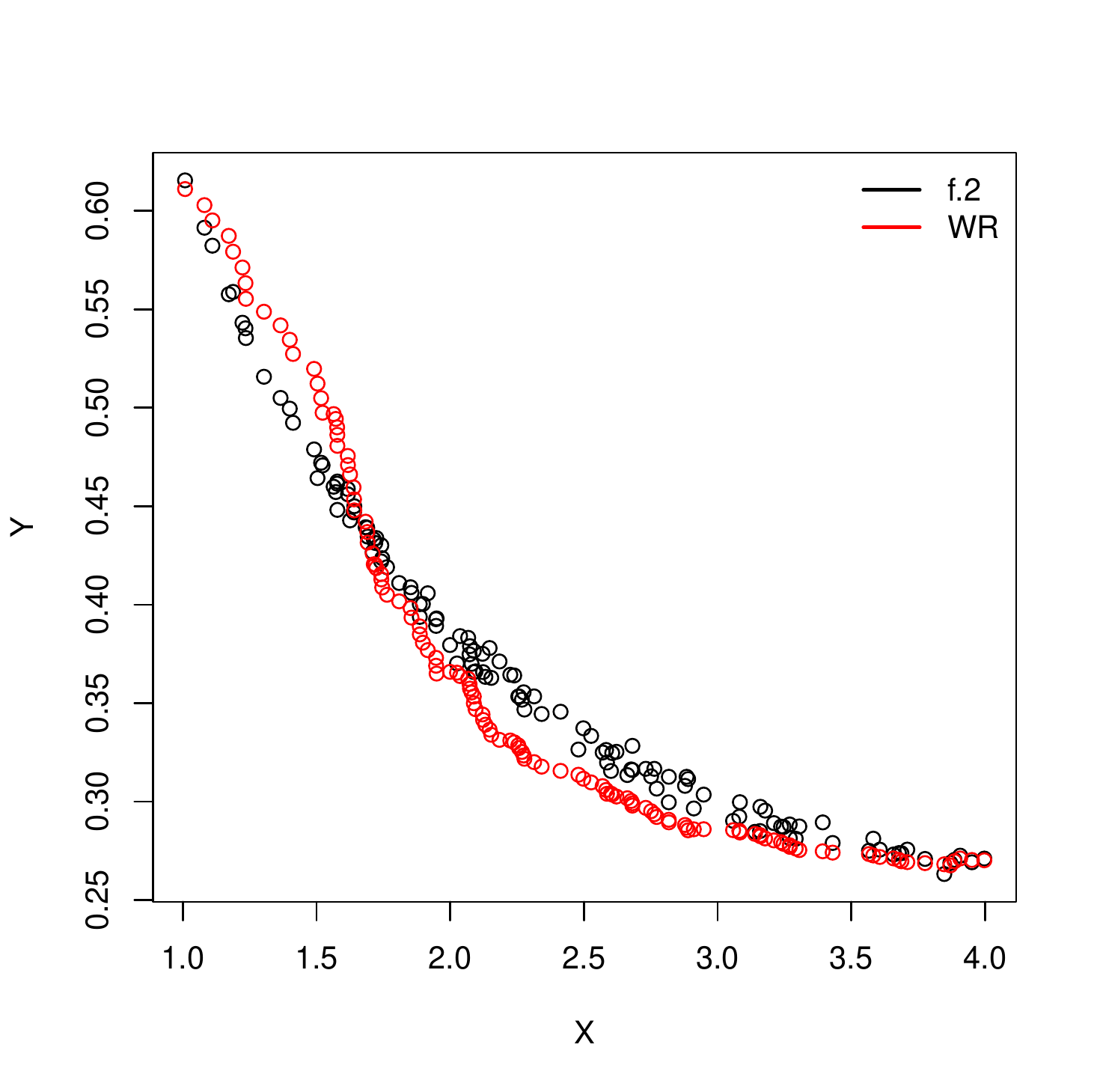}\\
	\includegraphics[width=5cm]{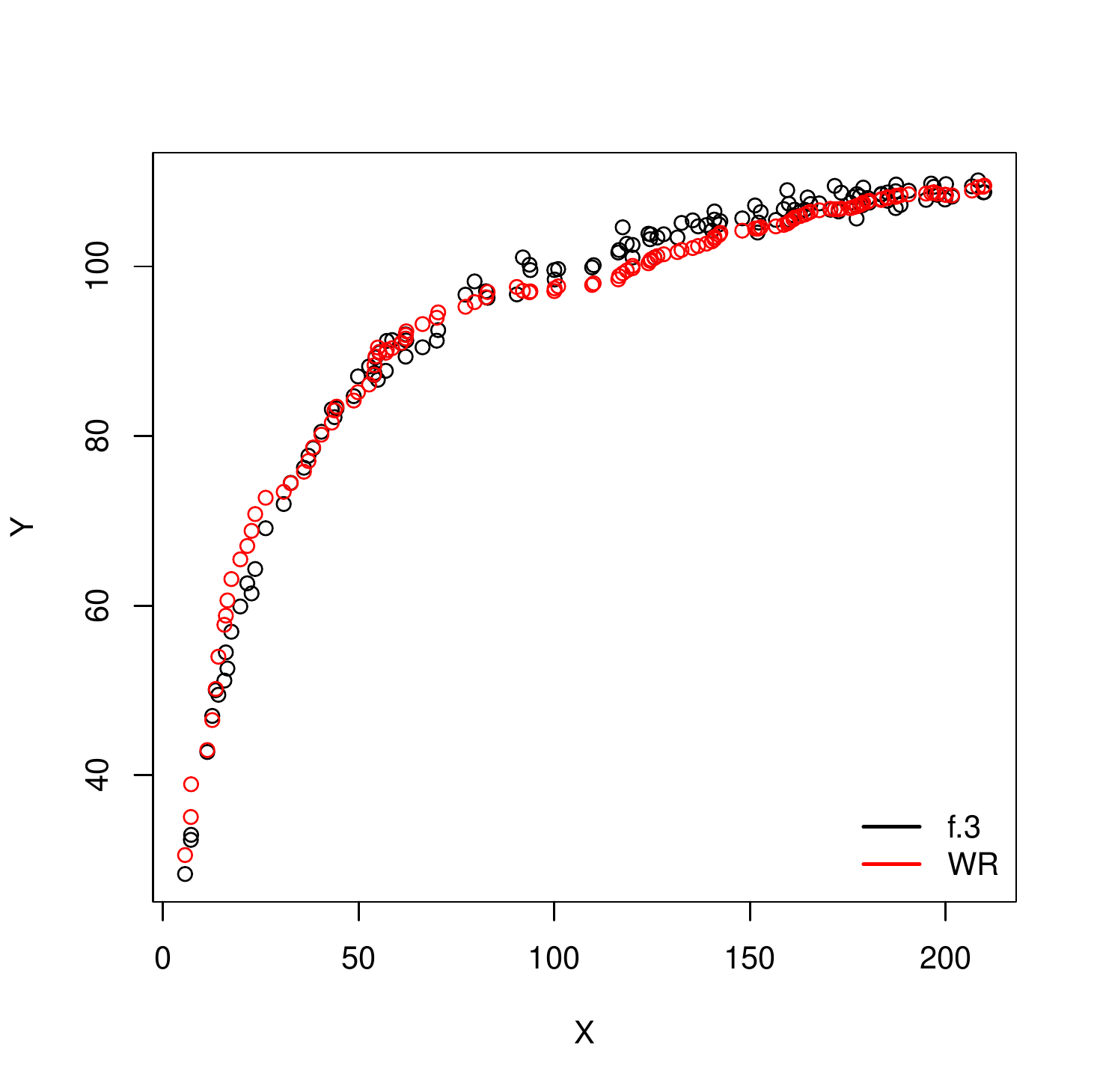}\includegraphics[width=5cm]{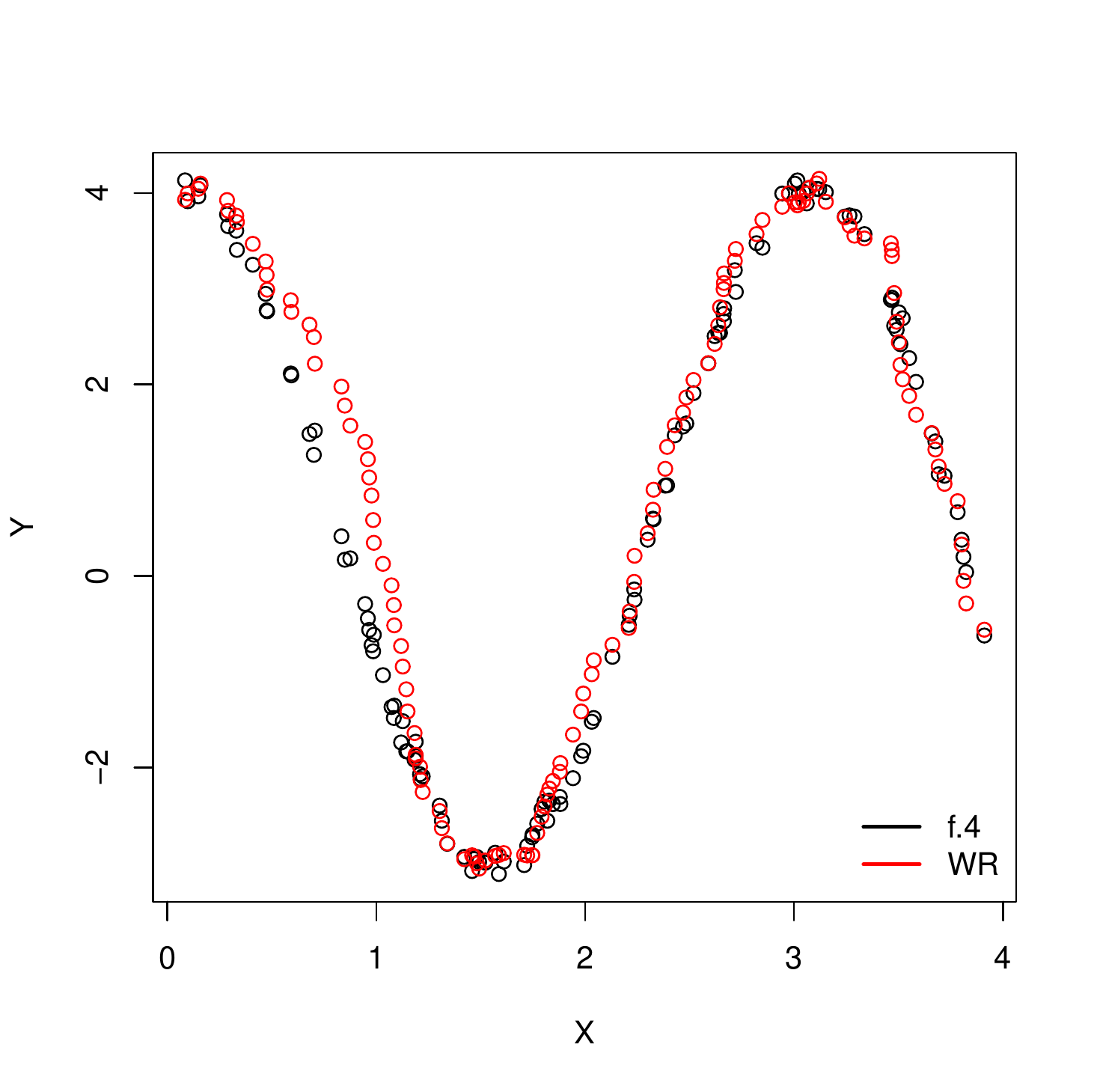}
\end{figure}

\textcolor{black}{The synthetic data sets are generated according to 36 different configurations taking into account: 3 sample sizes (128, 256, 512), 4 true nonlinear regression models ($f_1$, $f_2$, $f_3$, $f_4$) and 3 dependence levels (weak, moderate, strong) for the relationship between $Y$ and $X$. The explanatory variable $X$ is uniformly distributed in a predefined interval and the error terms $\epsilon$ signals the dependence level between $Y$ and $X$. Table \ref{parameters} brings the setting parameters considered in the scenario 1. Figure \ref{figuras} illustrates the nonlinear regression models (\ref{medicine})-({\ref{sincos}}) with a sample size $n=128$ and with a strong dependence degree between $Y$ and $X$. The red dots represents the fitted values for the wavelet non-parametric model.}
\begin{table}[htb]
	\centering
	\caption{Setting parameters and true nonlinear regression models for scenario 1. }\label{parameters}
	\resizebox{14cm}{!}{
		\begin{tabular}{c|c|ccc|ccc}
			\hline 
			\multirow{2}{*}{True model} & \multirow{2}{*}{$X$} & \multicolumn{3}{c|}{Dependence level}& \multicolumn{3}{c}{vetor $\beta$}  \\ \cline{3-8}
			&  & Strong & Moderate & weak &$\beta_1$ & $\beta_2$ & $\beta_3$ \\
			\hline
			$f_1$ & $X\sim U(-6,6)$  & $\varepsilon \sim N(0,0.01)$ & $\varepsilon \sim N(0,0.1)$ & $\varepsilon \sim N(0,0.2)$ & $2.00$ & $3.00$ & $1.00$ \\
			$f_2$ & $X\sim U(1,4)$  & $\varepsilon \sim N(0,0.005)$ & $\varepsilon \sim N(0,0.03)$ & $\varepsilon \sim N(0,0.06)$  & $0.25$ & $1.00$ & $-$ \\
			$f_3$ & $X\sim U(5,210)$  & $\varepsilon \sim N(0,1)$ & $\varepsilon \sim N(0,5)$ & $\varepsilon \sim N(0,10)$ & $20.0$ & $120.0$ & $-$ \\
			$f_4$ & $X\sim U(0,4)$  & $\varepsilon \sim N(0,0.1)$ & $\varepsilon \sim N(0,1)$ & $\varepsilon \sim N(0,2)$ & $4.00$ & $1.00$ & $-$ \\ 
			\hline
		\end{tabular} 
	}
\end{table}

\textcolor{black}{We considered a Monte Carlo simulation with 1,000 replications for each configuration.  At each time, we generate an artificial data set according to a predefined true nonlinear regression model. Then, the wavelet non-parametric model and other 24 eligible models are fitted to the data. The distance between the fitted values of the wavelet model and fitted values (including the true model) for each one of the candidate nonlinear models is obtained according to the following performance measures: root mean square error (RMSE) and median absolute error (MAE). If the minimal distance occurs between the true nonlinear model and the wavelet regression, we consider that the wavelet procedure (WP) presented a true classification. Finally, we evaluate the performance of the WP in terms of the percentage of true classification for each criterion (RMSE and MAE).}

\textcolor{black}{The Table \ref{tab:trueclasnonlinear} exhibits the true classification rate for the WP according with the true nonlinear regression model, sample size and criterion. The WP presented a percentage of true classification equal to 100\% for the model $f_1$ with moderate/strong dependence level between $Y$ and $X$. When the dependence is weak and the sample size $n=128$, the WP presented a low true classification rate in both criteria ($RMSE = 7.0\%$ and $MAE = 0.0\%$). This means that a false model was wrongly selected by the WP for this configuration. However, when $n=256$ or higher, the WP identified the true model 100\% of times for $RMSE$ and 96.4\% of times for $MAE$.}

\textcolor{black}{In relation to the true function $f_2$, the results demonstrate that the WP presents a better performance in comparison with the model $f_1$. Notice that the true classification rate is always higher than 99.5\%, except when the dependence level is weak and $MAE$ criterion. The performance of the WP for the model $f_3$ also demonstrated a good true classification rate when the dependence level is moderate or strong. An atypical result was found when the dependence level is weak, $n=512$ and $RMSE$ criterion. For this setup the true classification rate was 0.00 \%. However, the $MAE$ criterion presented a true classification rate equal to 99.3\%. For the parametric form $f_4$ the WP demonstrated an unsatisfactory performance when the dependence level is weak and $n=128$. However, for $n=256$ or higher, the WP identified the true model 100\% in both criteria. The same behavior occurred when the dependence level is moderate or strong. Another atypical result was found when the dependence level is strong, $MAE$ criterion and $n=512$. However, the $RMSE$ criterion presented a true classification rate equal to 100.0\%. Overall, the wavelet procedure demonstrated a high true classification rate in detect the true nonlinear parametric model.}
\begin{table}[htb]
	\caption{Scenario 1 - percentage of true classification of the wavelet procedure (WP) according with the true nonlinear model, sample size and criterion.}
	{\footnotesize
		\begin{center}
			\begin{tabular}{c|c|ccc|ccc}
				\hline \multirow{3}{*}{True model} & \multirow{3}{*}{Dep. Level} &  \multicolumn{3}{c|}{$RMSE$} & \multicolumn{3}{c}{$MAE$} \\ \cline{3-8}
				&   & \multicolumn{3}{c|}{$n$} & \multicolumn{3}{c}{$n$} \\ 
				& 	 & 128  & 256  & 512   & 128 & 256  & 512   \\ \hline
				\multirow{3}{*}{$f_1$} & Weak  & 7.0 & 100.0 & 100.0 & 0.0 & 96.4 & 100.0  \\
				&   Moderate & 100.0 & 100.0 & 100.0 & 100.0 & 100.0 & 100.0 \\ 
				&   Strong & 100.0 & 100.0 & 100.0 & 100.0 & 100.0 & 100.0 \\ \hline
				\multirow{3}{*}{$f_2$} & Weak  & 99.7 & 99.7 & 99.6 & 68.6 & 99.8 & 99.9  \\
				&   Moderate & 99.9 & 100.0 & 100.0 & 99.6 & 99.8 & 99.8 \\ 
				&   Strong & 100.0 & 100.0 & 100.0 & 100.0 & 99.8 & 100.0 \\ \hline
				\multirow{4}{*}{$f_3$} & Weak  & 100.0 & 100.0 & 0.0 & 79.4 & 100.0 & 99.3  \\
				&   Moderate & 100.0 & 100.0 & 100.0 & 99.5 & 100.0 & 97.8 \\ 
				&   Strong & 100.0 & 100.0 & 100.0 & 100.0 & 100.0 & 100.0 \\ \hline 
				\multirow{4}{*}{$f_4$} & Weak  & 0.0 & 100.0 & 100.0 & 0.0 & 100.0 & 100.0  \\
				&   Moderate & 100.0 & 100.0 & 100.0 & 100.0 & 100.0 & 100.0 \\ 
				&   Strong & 100.0 & 100.0 & 100.0 & 100.0 & 100.0 & 0.0 \\ 
				\hline
			\end{tabular}
			\label{tab:trueclasnonlinear}
		\end{center}
	}
\end{table}

\subsection*{\textcolor{black}{Scenario 2: performance of the WP when the nonlinear functions are quite similar}}

\textcolor{black}{In the previous section we evaluated the WP to identify the true nonlinear function taking into account a wide range of candidate models. It is possible to believe that the WP procedure can to present a good performance when the candidate nonlinear models are quite different. To check this point we evaluate the performance of the WP when the nonlinear functions are quite similar or present a very similar behavior. 
	We considered the nonlinear function $f_2$ as the true model and to generate the synthetic data. The function $f_{24}$ was considered as competitor model, being defined by:}
\begin{equation}
y=f_{24}(x,\beta)=\frac{1}{\beta_1 + \beta_2 x} + \epsilon. \nonumber \label{f24}
\end{equation}

\begin{figure}[htb]
	\centering
	\caption{Empirical relation between $Y$ and $X$ according to the nonlinear function $f_2$. Fitted values for the candidate model $f_{24}$ and wavelet regression, according to the sample size and dependence level: weak (A), moderate (B) and strong (C).}\label{fsemelhantes}
	{$n=128$}\\
	\includegraphics[width=4cm]{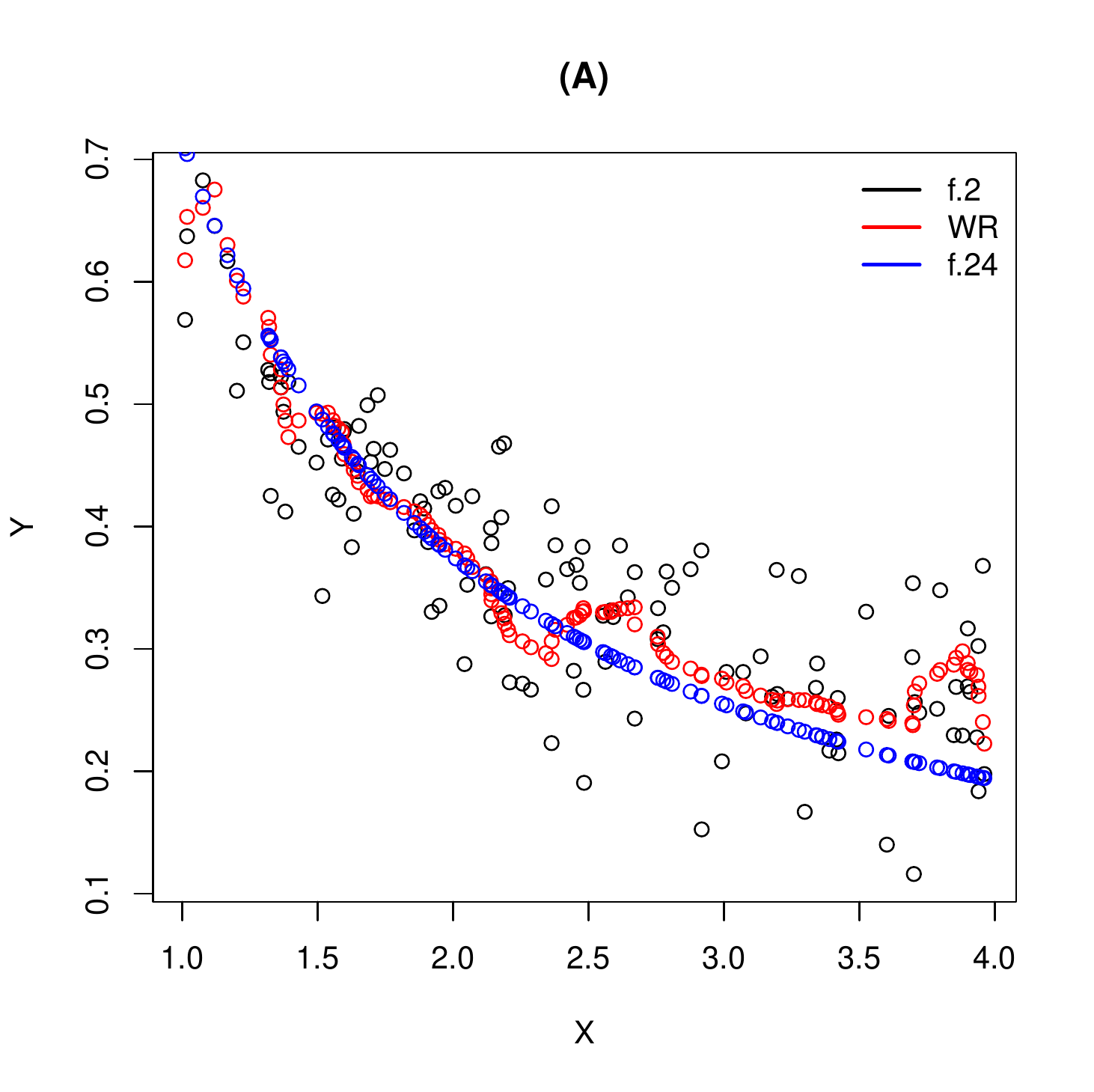}\includegraphics[width=4cm]{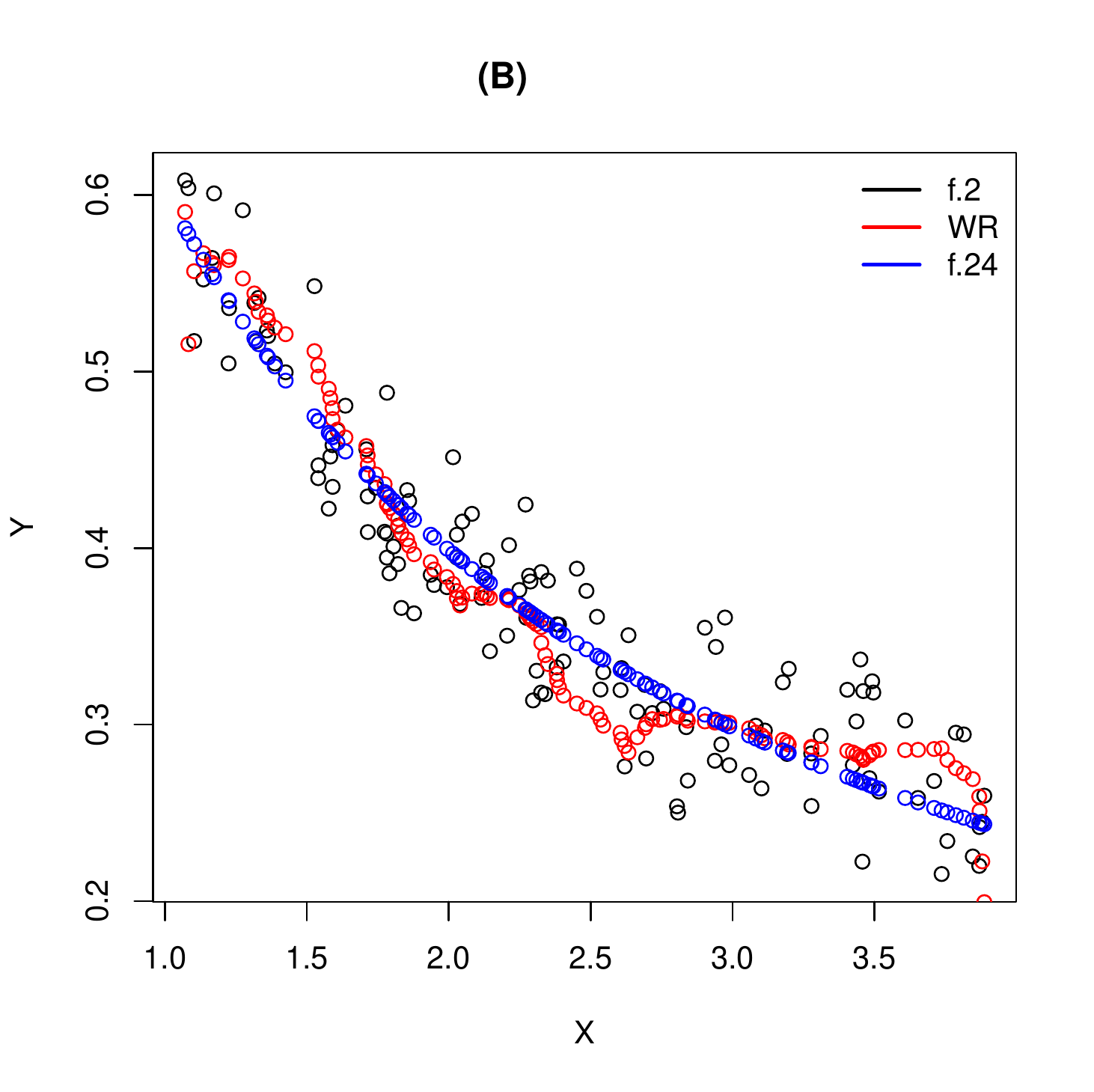}\includegraphics[width=4cm]{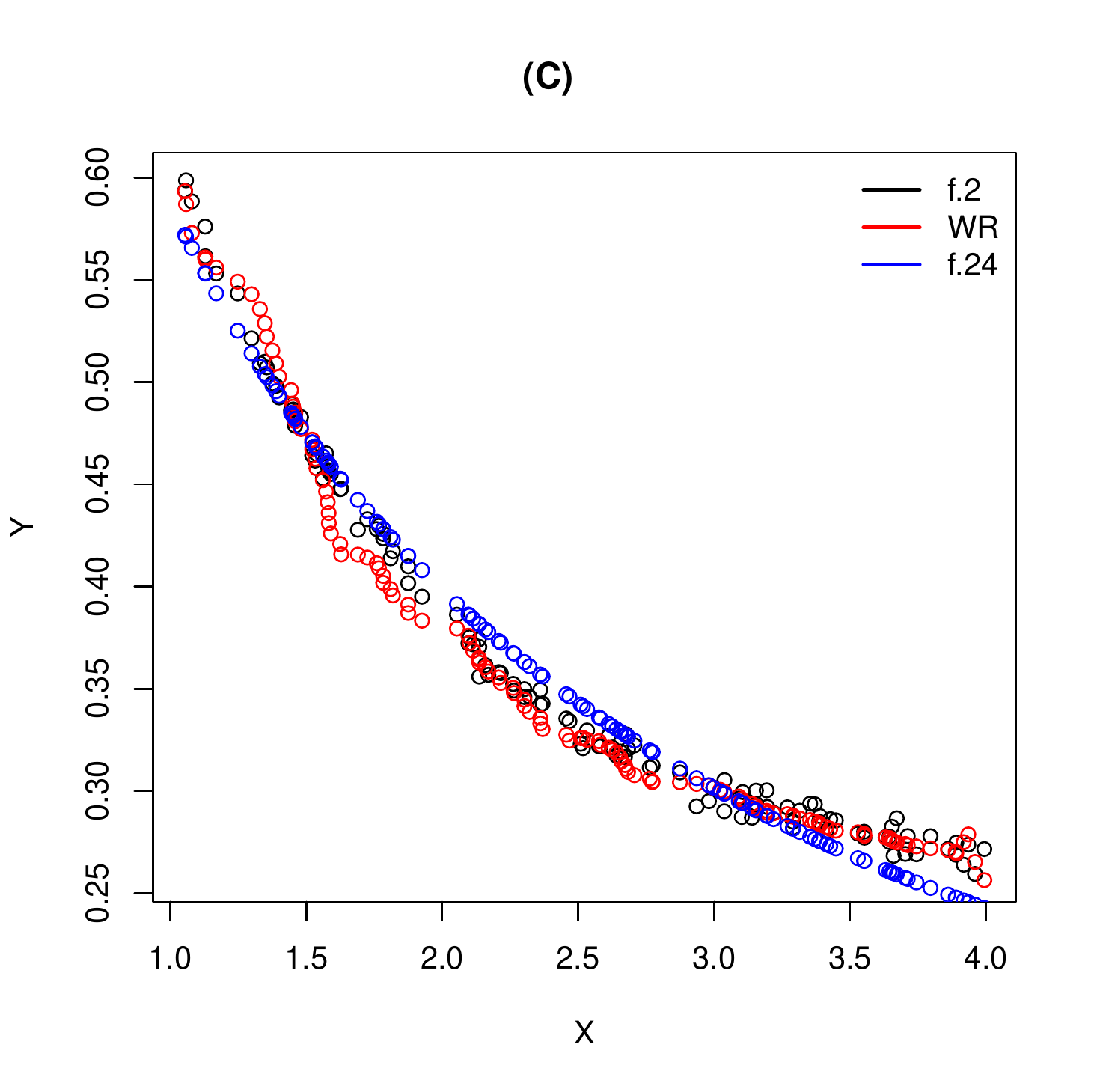} \\
	{$n=256$}\\
	\includegraphics[width=4cm]{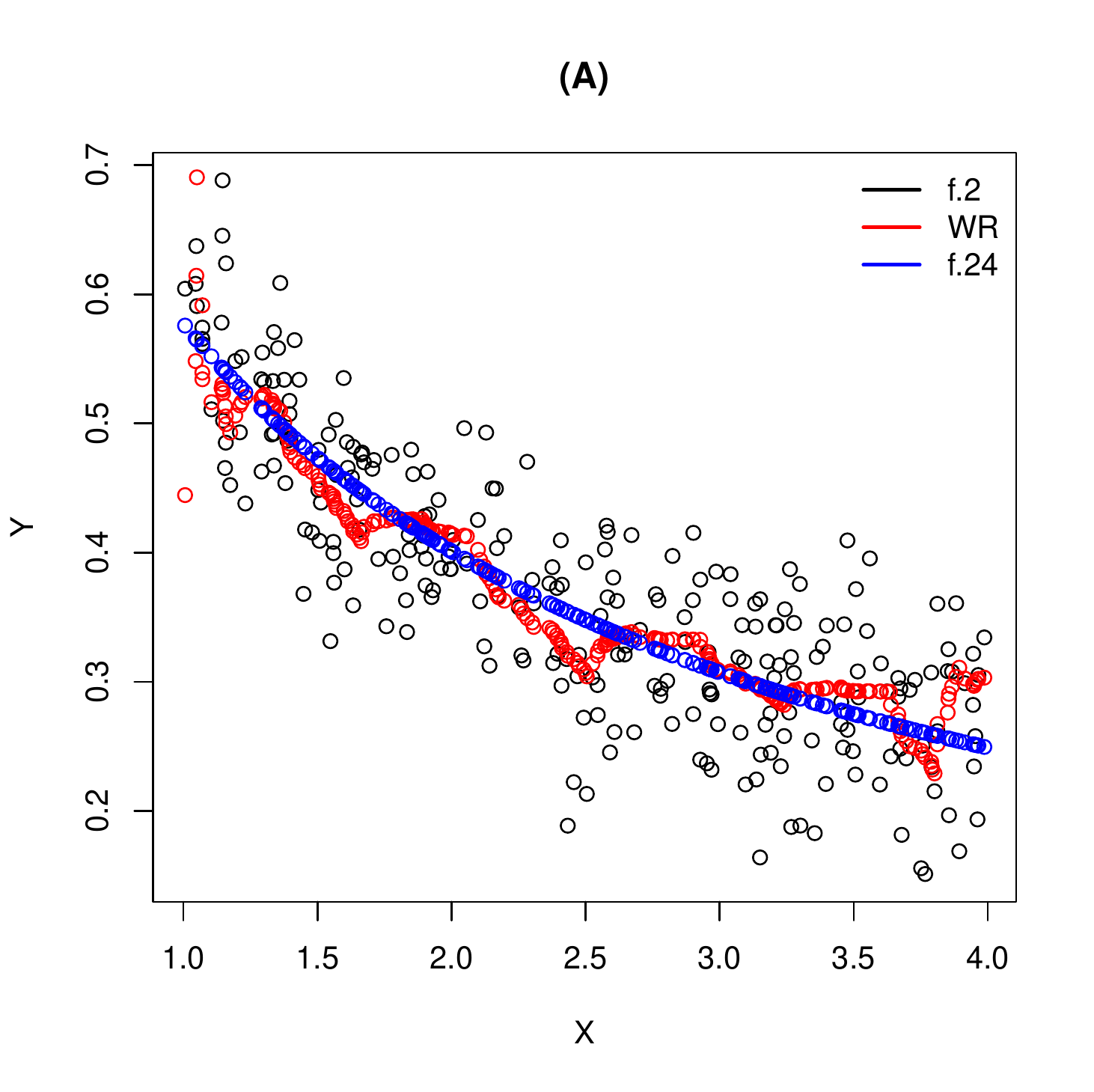}\includegraphics[width=4cm]{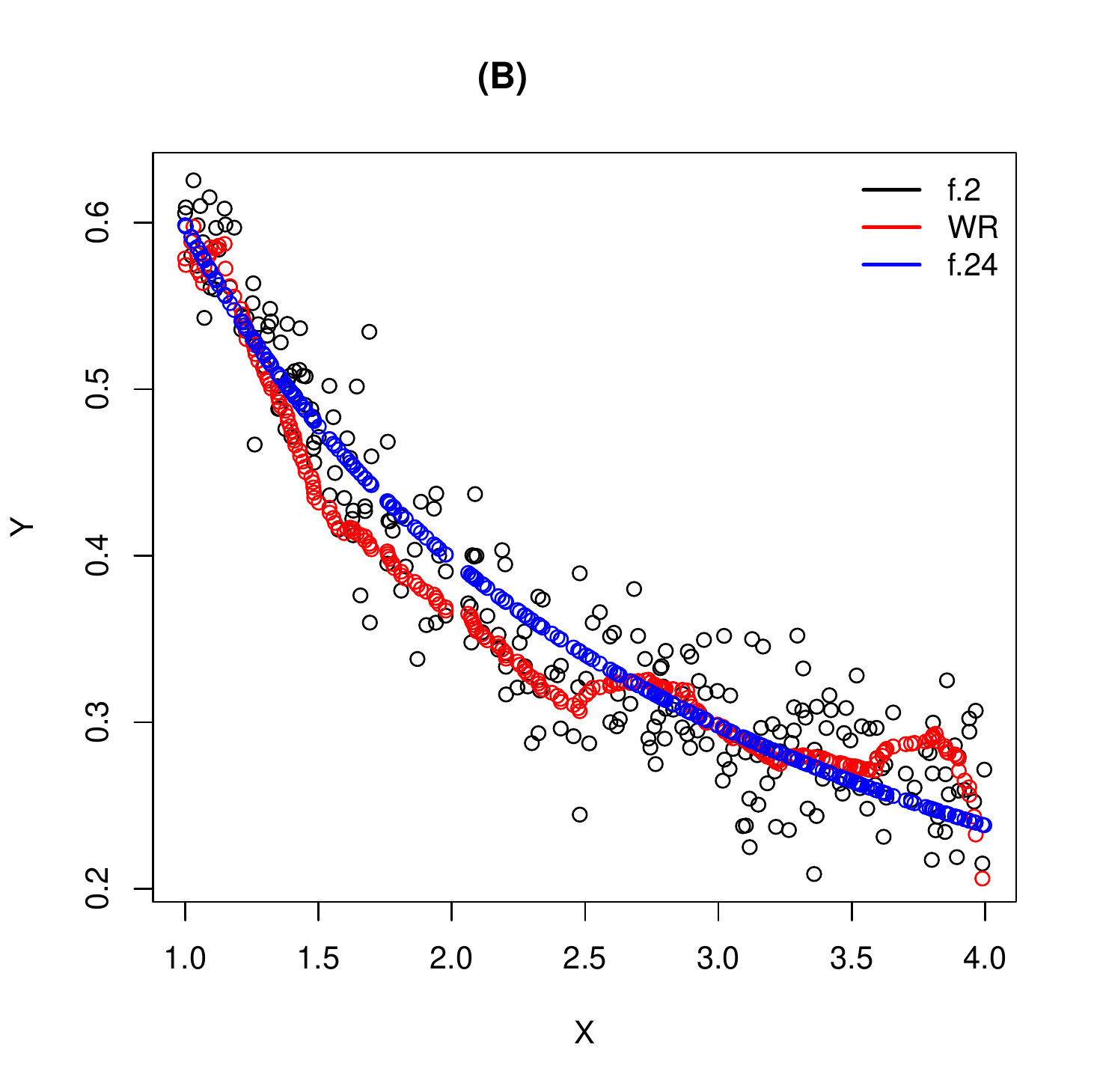}\includegraphics[width=4cm]{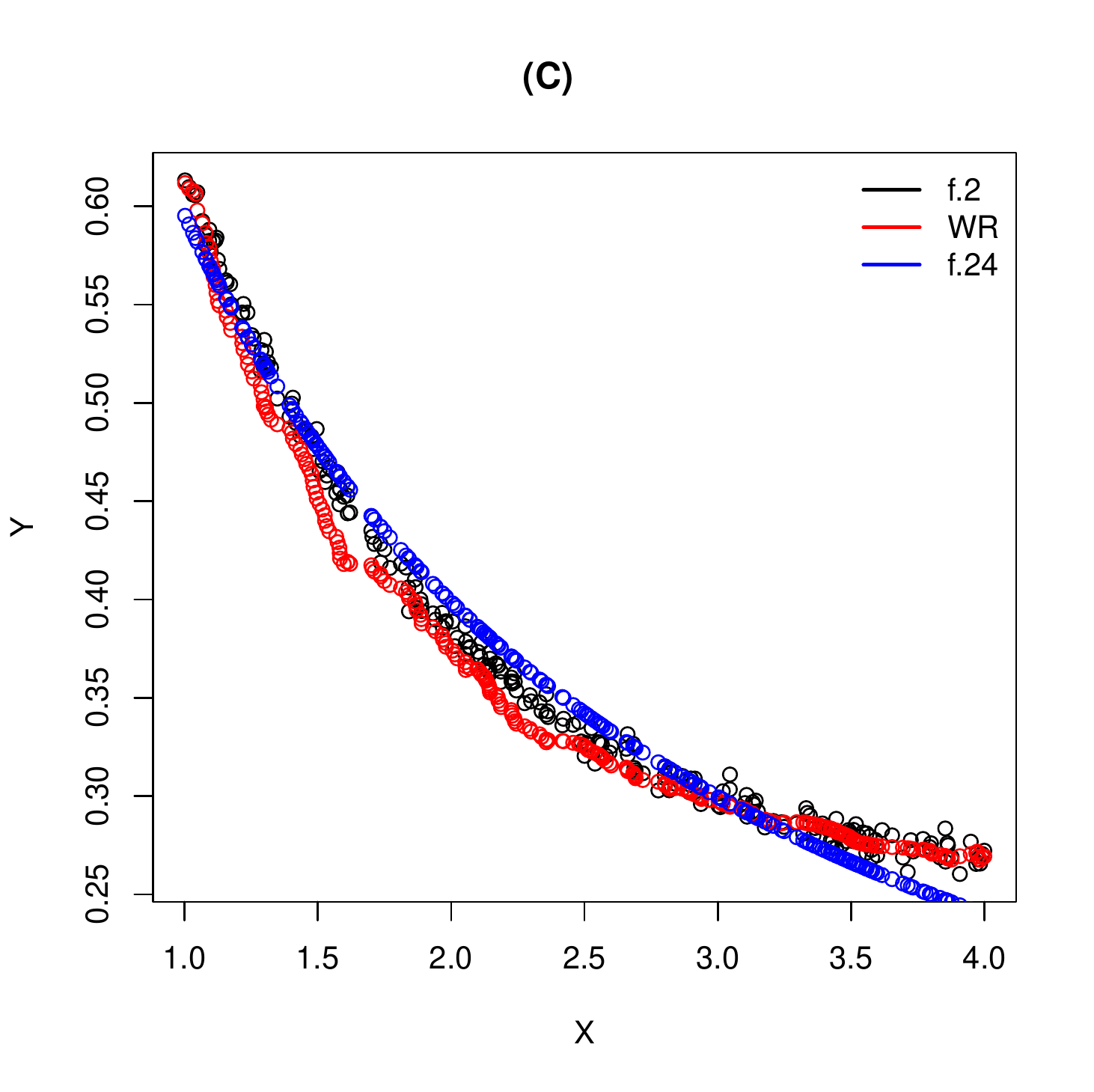} \\
	{$n=512$}\\
	\includegraphics[width=4cm]{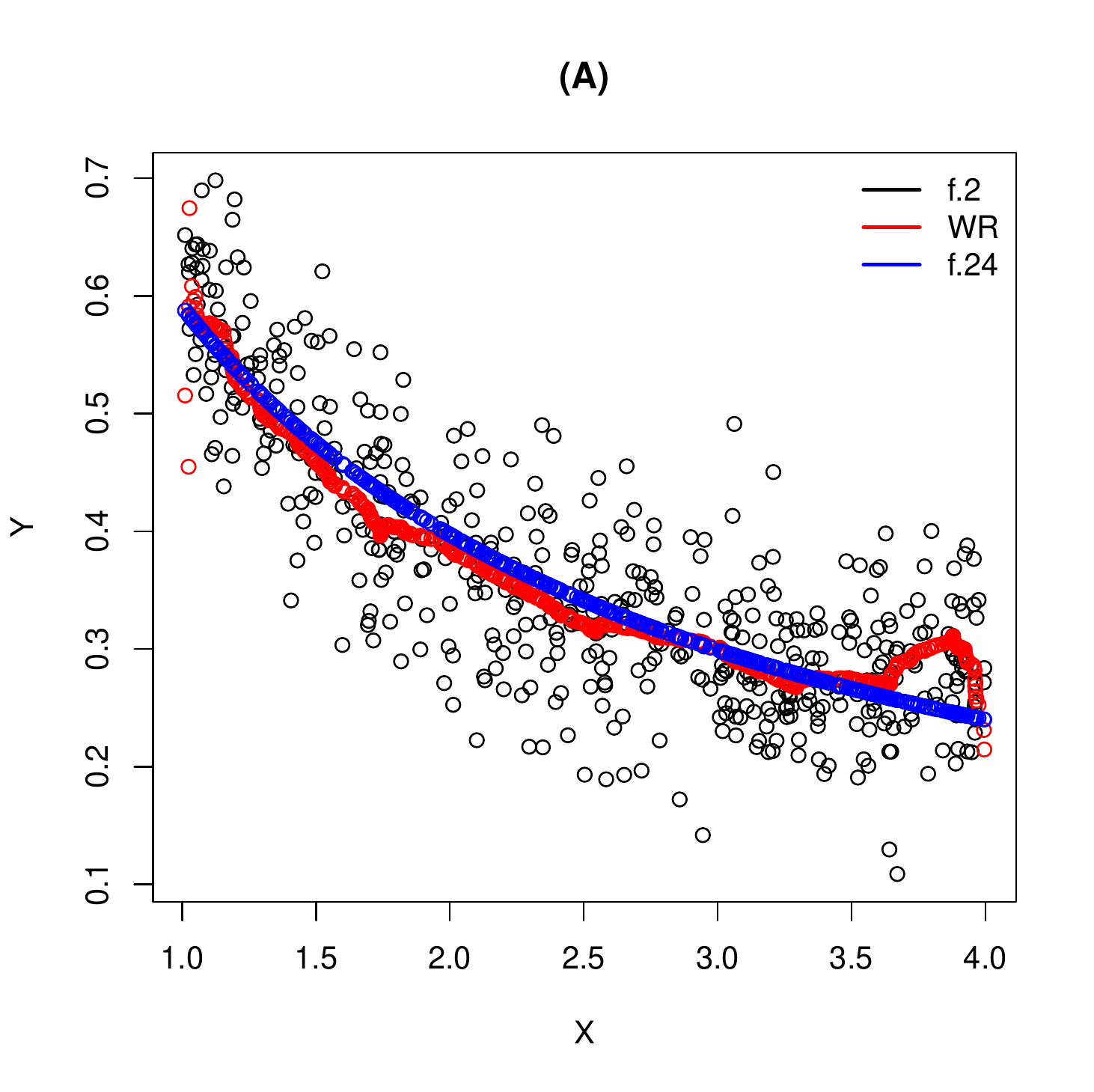}\includegraphics[width=4cm]{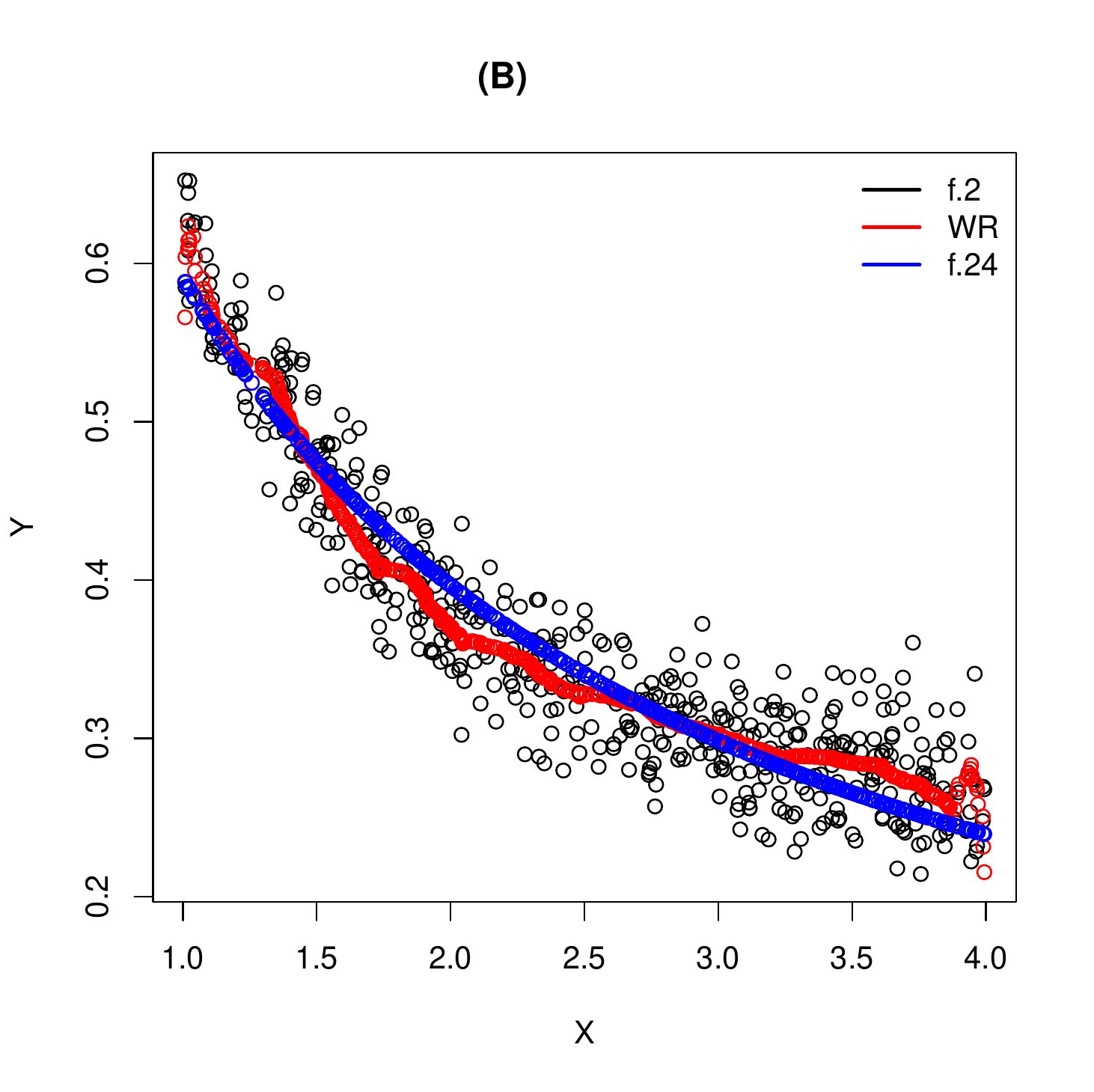}\includegraphics[width=4cm]{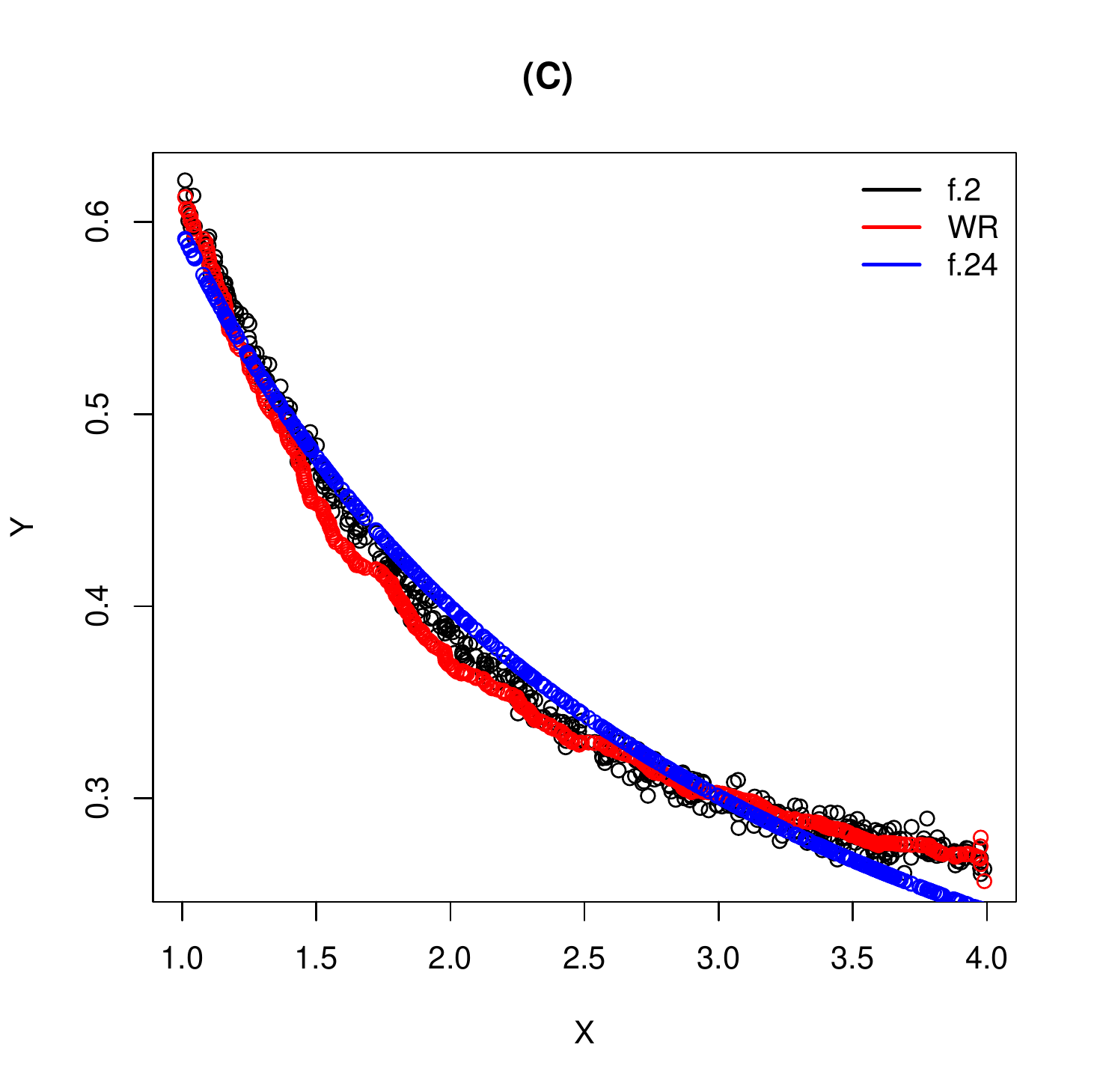}
\end{figure}

\textcolor{black}{Figure \ref{fsemelhantes} illustrates the behavior of the synthetic data based on the true function $f_2$ (black points). The blue points represent the fitted values for the competitor model ($f_{24}$) while the red points represent the fitted values for the wavelet model. We considered a total of 12 different configurations. Notice that when the dependence level between $Y$ and $X$ is weak or moderate, the wavelet model seems to be more sensible to the noise existing in the data. However, when the dependence level is strong the competitor model presented a lack of fit at the bottom of the data.}

\textcolor{black}{We considered a Monte Carlo simulation with 1,000 replicates for each configuration. At each replication, we generate an artificial data set according with the true nonlinear regression model. Then, the wavelet non-parametric model and the nonlinear functions ($f_2$ and $f_{24}$) are fitted to the data. Finally, the distances between the fitted values of the wavelet model and fitted values of the parametric models (including the true model) are obtained according to the performance measures RMSE and MAE.}

\textcolor{black}{Table \ref{tf2e24} presents the true classification rate for the WP based on the Monte Carlo experiments. The results suggest that the WP presents a good true classification rate also in this scenario. This means that the WP identifies the true nonlinear model even when the competitor model presents a very similar behavior.}
\begin{table}[htb]
	\centering
	\caption{Percentage of true classification for the WP. Comparative study between the true model $f_2$ and the competitor model $f_{24}$.}\label{tf2e24}
	\begin{tabular}{c|cc|cc|cc}
		\hline
		& \multicolumn{2}{c|}{Weak} & \multicolumn{2}{c|}{Moderate} & \multicolumn{2}{c}{Strong} \\
		$n$ & RMSE & MAE & RMSE & MAE & RMSE & MAE \\
		\hline
		128 & 100.0 & 91.8 & 100.0 & 100.0 & 100.0 &100.0\\ 
		256 & 100.0 & 100.0 & 100.0 & 100.0 & 100.0 & 100.0\\ 
		512 & 100.0 & 100.0 & 100.0 & 100.0 & 100.0 & 100.0\\ 
		\hline
	\end{tabular}
\end{table}

\subsection*{\textcolor{black}{Scenario 3: identifying the more appropriate link function for a GLM model}}
Now, we perform an experimental study to evaluate the performance of the WP to identify the true link function for a GLM model. The artificial data sets consider a predefined relationship (true link function) between the response variable $Y$ and the linear predictor $\bm{\eta}=\bm{X\beta}$, where \textbf{X} represents the matrix model. The synthetic data sets are generated according to 30 different configurations, taking into account 3 sample sizes (128, 256, 512), 3 probability distributions for the response variable $Y$ (Gaussian, gamma, inverse Gaussian) and 4 link functions (identity, logarithm, inverse, $1/\mu^2$). The link function $1/\mu^2$ was considered only for the inverse Gaussian model. We considered one explanatory variable $X$, uniformly distributed in the interval $[0.5,1.5]$. 

The synthetic data sets are built considering a Monte Carlo simulation with 1,000 replications for each configuration.  At each time, we generate an artificial data set according to a predefined GLM. The wavelet model and the other eligible GLMs are fitted to the synthetic data set. Then, we compute the distances between the fitted values of the wavelet model and the others eligible GLM's (including the true model) according to the performance measures RMSE and MAE. If the minimal distance, between the wavelet model and the eligible GLM's, occurs for the true GLM, we consider that the WP presented a true classification. 

Table \ref{tab:trueclas} exhibits the percentage of the true classification for the WP. We verify that the true classification rate for the RMSE criterion presented higher values than the MAE criterion. Thus, the use of the RMSE criterion will be preferable to decide the more appropriate link function. Moreover, the results demonstrated that the WP was efficient tool to identify the appropriate link function for a GLM. Notice that the percentage of true classification increases when the sample size increases. If we consider the RMSE criterion, the true classification rate is higher than 80\% in the majority of the configurations. The {\it log} and {\it inverse} link functions exhibited the best true classification rate, when compared with the {\it identity} link. The WP also demonstrated a better performance for the asymmetric distributions (gamma and inverse Gaussian), when compared with the Gaussian distribution. These results highlight that the WP can be used for choosing the appropriate link function when the response variable presents an asymmetric distribution and/or a nonlinear relationship between the variables.
\begin{table}[htb]
	\caption{Scenario 3 - true classification rate for the WP according to the random component, true link function, sample size and criterion.}
	{\footnotesize
		\begin{center}
			\begin{tabular}{c|c|ccc|ccc}
				\hline \multirow{3}{*}{Random Comp.} & \multirow{3}{*}{True link function} &  \multicolumn{3}{c|}{$RMSE$} & \multicolumn{3}{c}{$MAE$} \\ \cline{3-8}
				&   & \multicolumn{3}{c|}{$n$} & \multicolumn{3}{c}{$n$} \\ 
				& 	 & 128  & 256  & 512   & 128 & 256  & 512   \\ \hline
				\multirow{3}{*}{Gaussian} & Identity  & 85.7 & 94.2 & 98.4 & 79.8 & 85.0 & 88.4  \\
				&   inverse & 91.7 & 97.5 & 99.7 & 89.6 & 93.7 & 95.8 \\ 
				&   log & 99.4 & 100.0 & 100.0 & 98.0 & 99.6 & 100.0 \\ \hline
				\multirow{3}{*}{Gama} & Identity  & 88.1 & 95.3 & 99.0 & 79.0 & 82.2 & 89.2  \\
				&   inverse & 91.3 & 97.9 & 100.0 & 80.0 & 84.7 & 92.1 \\ 
				&   log & 100.0 & 100.0 & 100.0 & 91.7 & 98.0 & 99.7 \\ \hline
				\multirow{4}{*}{Inverse Gaussian} & Identity  & 97.7 & 99.0 & 100.0 & 77.6 & 84.7 & 91.3  \\
				&   inverse & 95.7 & 99.2 & 100.0 & 53.1 & 70.4 & 84.7 \\ 
				&   log & 100.0 & 100.0 & 100.0 & 89.3 & 95.4 & 99.1 \\ 
				&   $1/\mu^2$ & 73.3 & 85.9 & 93.7 & 63.8 & 75.1 & 80.6 \\ 
				\hline
			\end{tabular}
			\label{tab:trueclas}
		\end{center}
	}
\end{table}

Additionally, we evaluate the number of non null coefficients of the wavelet regression. This aspect allows identify if the non-parametric model is (or not) overfitting the data.  This was verified in terms of the percentage of null coefficients, after thresholding, by level. Table \ref{tab:nonull128} presents the percentage of null coefficients for the wavelet model, by level, after the thresholding ($n=128$). Note that the percentage of null coefficients is zero until level 2, for all configurations. After level 2, the results demonstrated that the majority of the coefficients of the wavelet  model are null. This means that the wavelet model requires few coefficients to fit to the data. Moreover,  this results signal that the wavelet model detects the more appropriate link function based on few non null coefficients, suggesting that the non-parametric model does not overfitting the data. We have obtained similar results for the sample sizes 256 and 512.
\begin{table}[htb]
	\caption{Percentage of null coefficients in wavelet model after threshold, according to level, random component and true link function ($n=128$).}
	{\footnotesize
		\begin{center}
			\begin{tabular}{c|c|ccc|ccc}
				\hline \multirow{2}{*}{Random Comp.} & \multirow{2}{*}{Link function} &  \multicolumn{6}{c}{$Level$} \\ \cline{3-8}
				& 	 & 1  & 2  & 3   & 4 & 5  & 6   \\ \hline
				\multirow{3}{*}{Gaussian} & Identity  & 0.00 & 0.00 & 74.8 & 87.3 & 93.7 & 96.2  \\
				&   inverse & 0.00 & 0.00 & 74.3 & 87.9 & 96.0 & 96.3 \\ 
				&   log & 0.00 & 0.00 & 72.0 & 87.3 & 94.2 & 96.3  \\ \hline
				\multirow{3}{*}{Gama} & Identity  & 0.00 & 0.00 & 74.9 & 87.4 & 93.7 & 96.2  \\
				&   inverse & 0.00 & 0.00 & 71.1 & 87.3 & 93.6 & 96.3  \\ 
				&   log & 0.00 & 0.00 & 73.1 & 87.4 & 94.7 & 96.2  \\ \hline
				\multirow{4}{*}{Inverse Gaussian} & Identity  & 0.00 & 0.00 & 76.4 & 89.6 & 94.8 & 94.9   \\
				&   inverse & 0.00 & 0.00 & 72.3 & 86.2 & 94.3 & 94.0  \\ 
				&   log & 0.00 & 0.00 & 64.8 & 76.0 & 81.4 & 81.4 \\ 
				&   $1/\mu^2$ & 0.00 & 0.00 & 74.5 & 89.4 & 96.2 & 95.8  \\ 
				\hline
			\end{tabular}
			\label{tab:nonull128}
		\end{center}
	}
\end{table}

\subsection*{\textcolor{black}{Scenario 4: comparing the predictive performance of the wavelet model against the GLM}}
\begin{figure}
	\centering
	\includegraphics[width=0.5\linewidth]{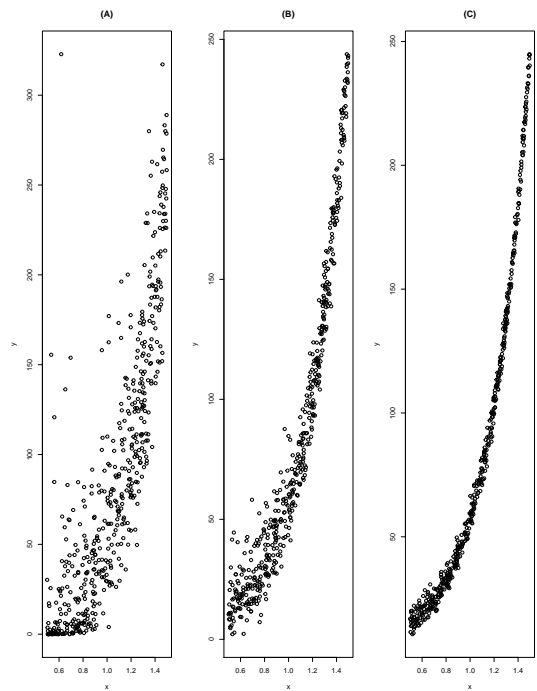}
	\caption{Illustrative scatter plot of $Y$ vs $X$. Gamma model with log link function. Dependence level weak (A), moderate (B) and strong (C).}
	\label{fig:Fig_level_dependences_Gaussian_Log}
\end{figure}

\begin{figure}
	\centering
	\includegraphics[width=0.5\linewidth]{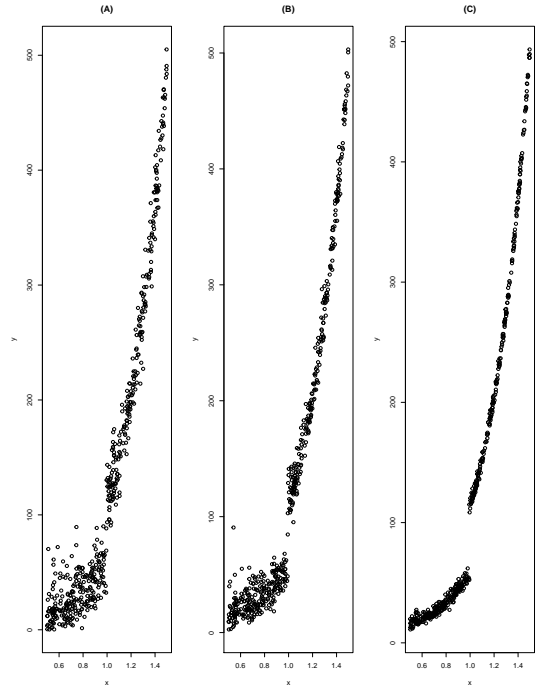}
	\caption{Illustrative scatter plot of $Y$ vs $X$. Gamma model with log link function (with gap). Dependence level weak (A), moderate (B) and strong (C).}
	\label{fig:Fig_level_dependences_Gaussian_Log_GAP}
\end{figure}

Based on the results presented in the previous scenarios, it is reasonable to ask if the wavelet model presents a better fitted values in comparison with the ``best'' parametric model. Thus, in the scenario 4, we evaluate the predictive performance of the wavelet regression against the true fitted GLM, taking into account three different sample and dependence levels between the response variable $Y$ and the linear predictor $\eta$.  Figure \ref{fig:Fig_level_dependences_Gaussian_Log} illustrates the dependence levels - weak (A), moderate (B) and strong (C) - for a gamma model with link function log. Usually, the parametric models present problems when the data exhibit this  characteristic that occurs due to a change in the slope parameter $\beta_0$. Thus, the predictive performance between the wavelet regression and the GLM it was also evaluated taking into account the presence of a gap in the data, as can be visualized in Figure \ref{fig:Fig_level_dependences_Gaussian_Log_GAP}.

A Monte Carlo simulation study with 1,000 replications was considered taking into account a predefined random component (gamma, Gaussian and inverse Gaussian) and a predefined regression structure between the mean of the response variable $Y$ and the linear prediction $\eta$. We also considered three different sample sizes $n=\{128, 256, 512\}$ and three dependence levels: weak (a), moderate (b) and strong (c). We also considered data with and without a gap, as illustrated in Figure \ref{fig:Fig_level_dependences_Gaussian_Log_GAP}, in a total of 90 different scenarios. Finally, for each configuration, the wavelet regression and GLM are fitted and the approaches are compared based on the performance measures RMSE and MAE. 

\begin{figure}
	\centering
	\subfloat[Dependence level weak]{\includegraphics[width=0.6\linewidth]{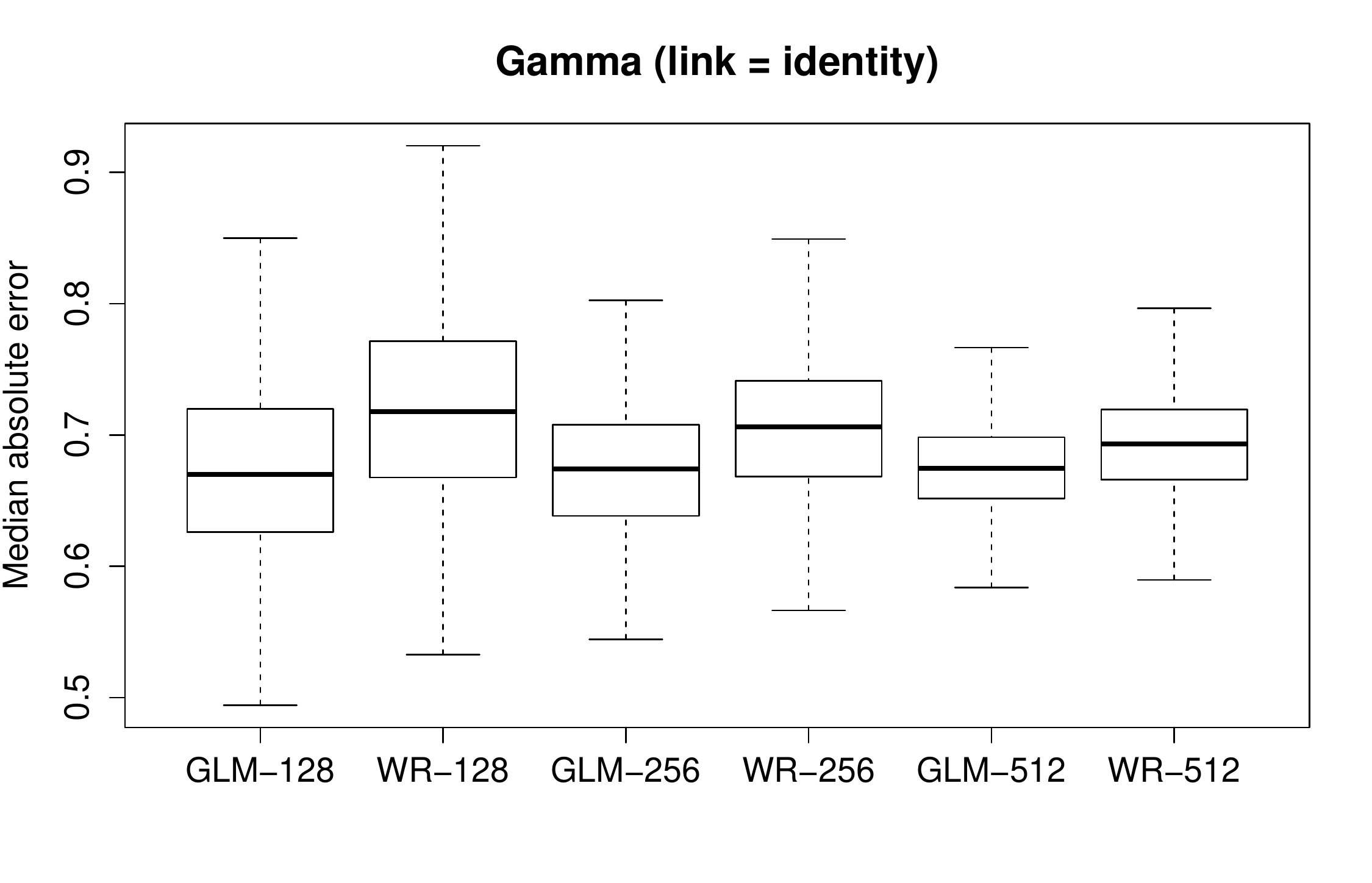}}\\
	\subfloat[Dependence level moderate]{\includegraphics[width=0.6\linewidth]{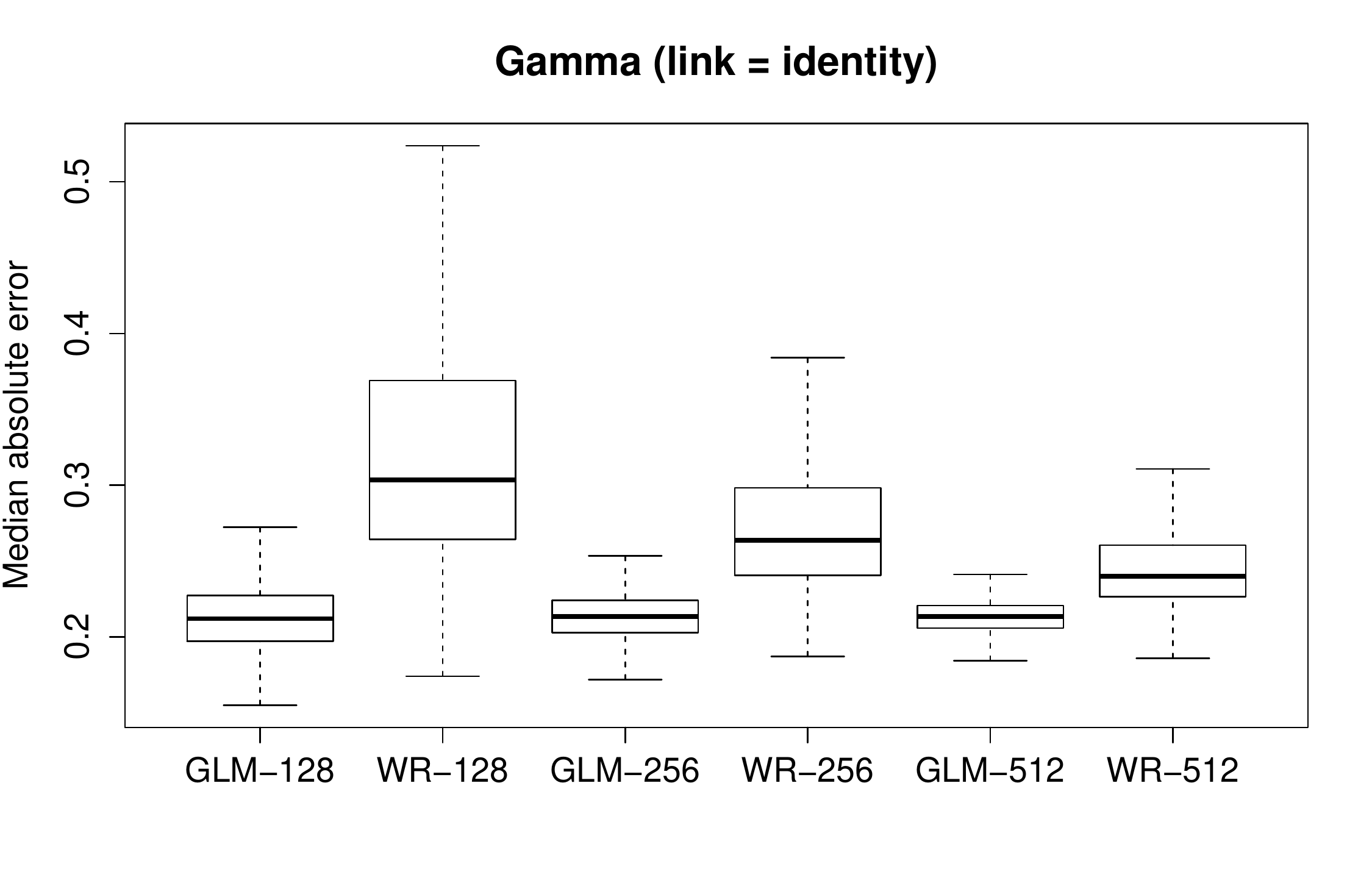}}\\
	\subfloat[Dependence level strong]{\includegraphics[width=0.6\linewidth]{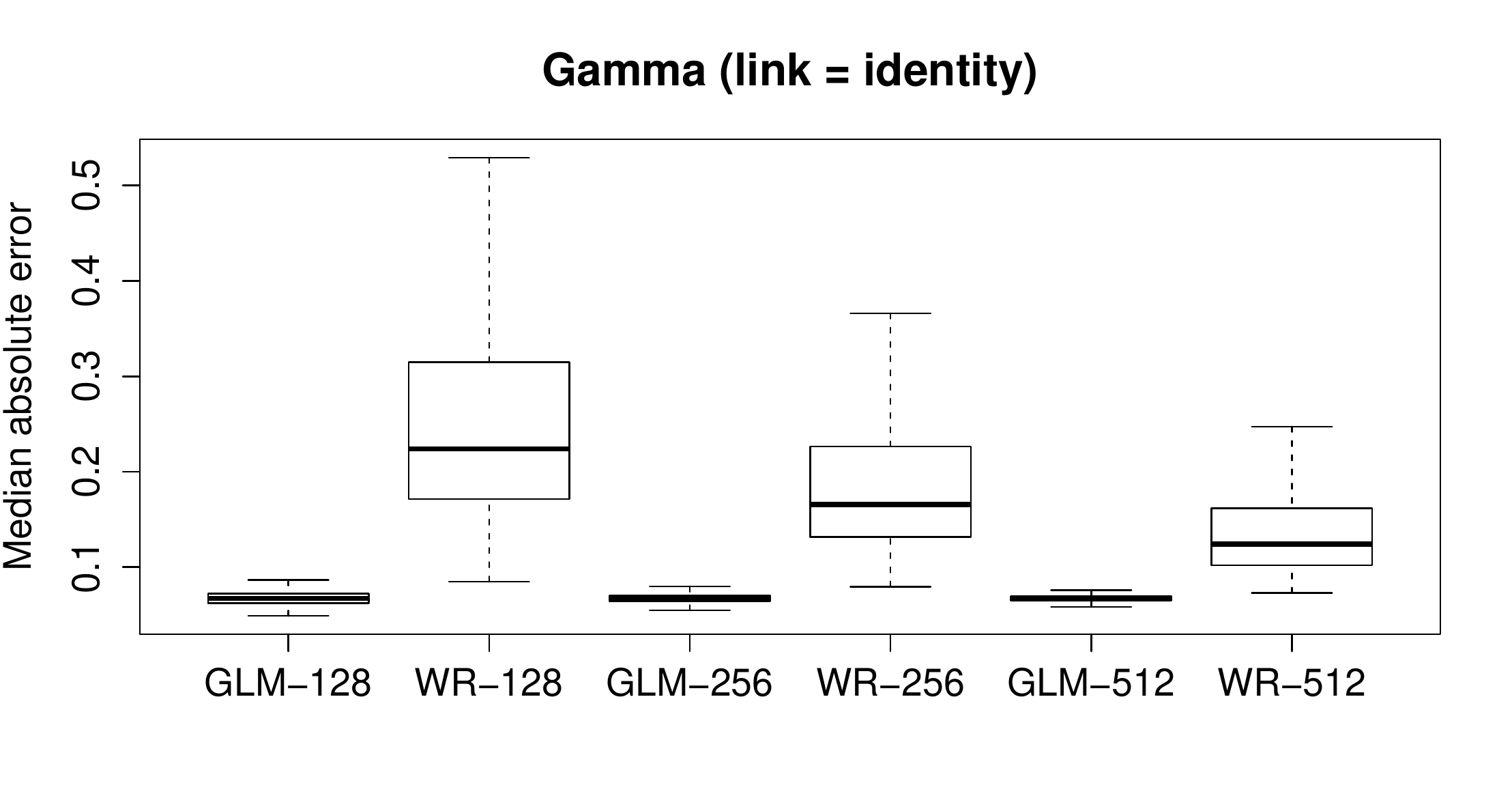}}
	\caption{Comparison between wavelet regression and GLM based on MAE.  Synthetic data sets with error gamma and link function identity link without gap. Dependence level weak (a), moderate (b) and strong (c).}
	\label{fig:Fig_Gama_identity_levels_MAE}
\end{figure}

\begin{figure}
	\centering
	\subfloat[Dependence level weak]{\includegraphics[width=0.6\linewidth]{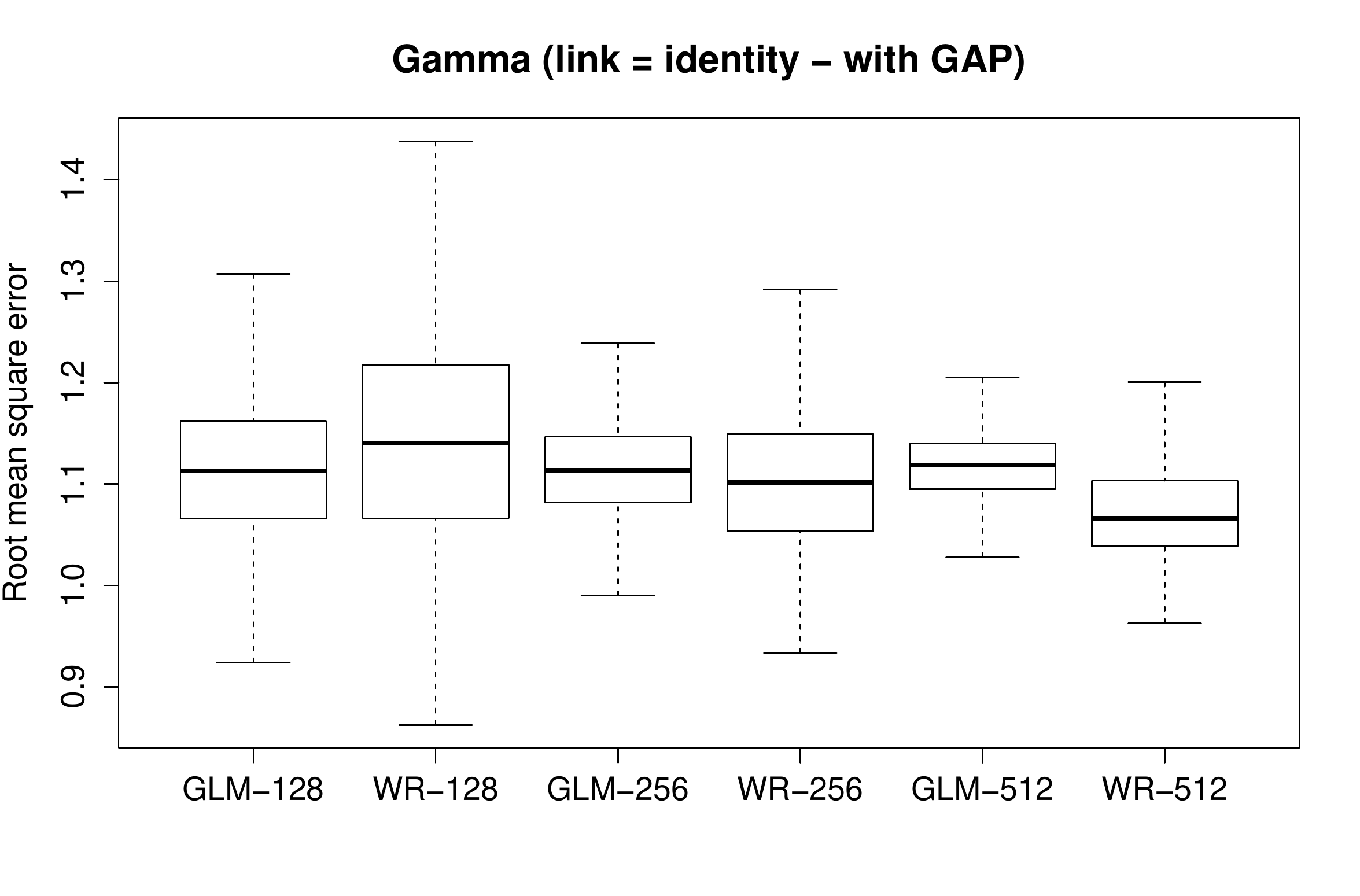}}\\
	\subfloat[Dependence level moderate]{\includegraphics[width=0.6\linewidth]{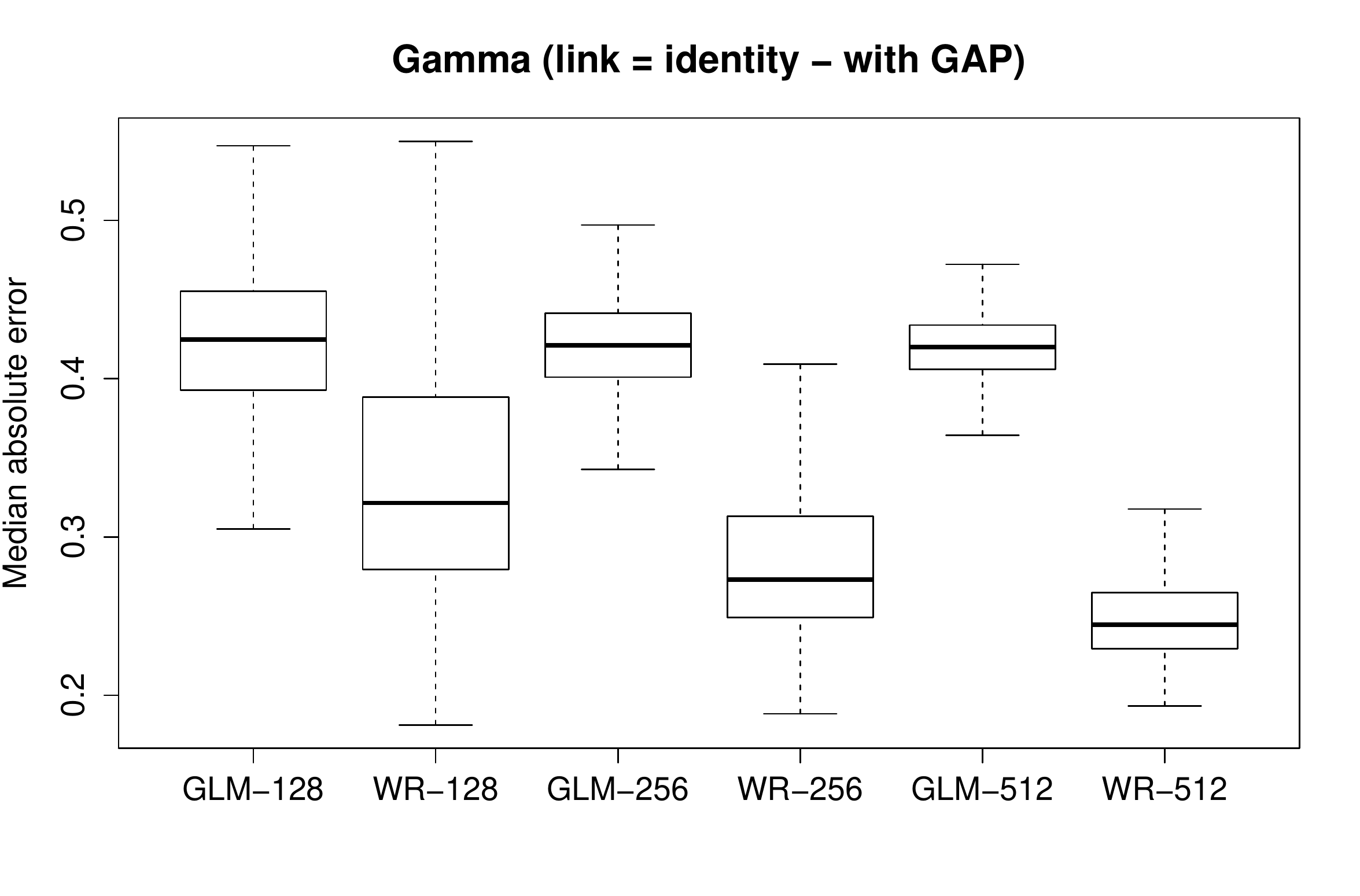}}\\
	\subfloat[Dependence level strong]{\includegraphics[width=0.6\linewidth]{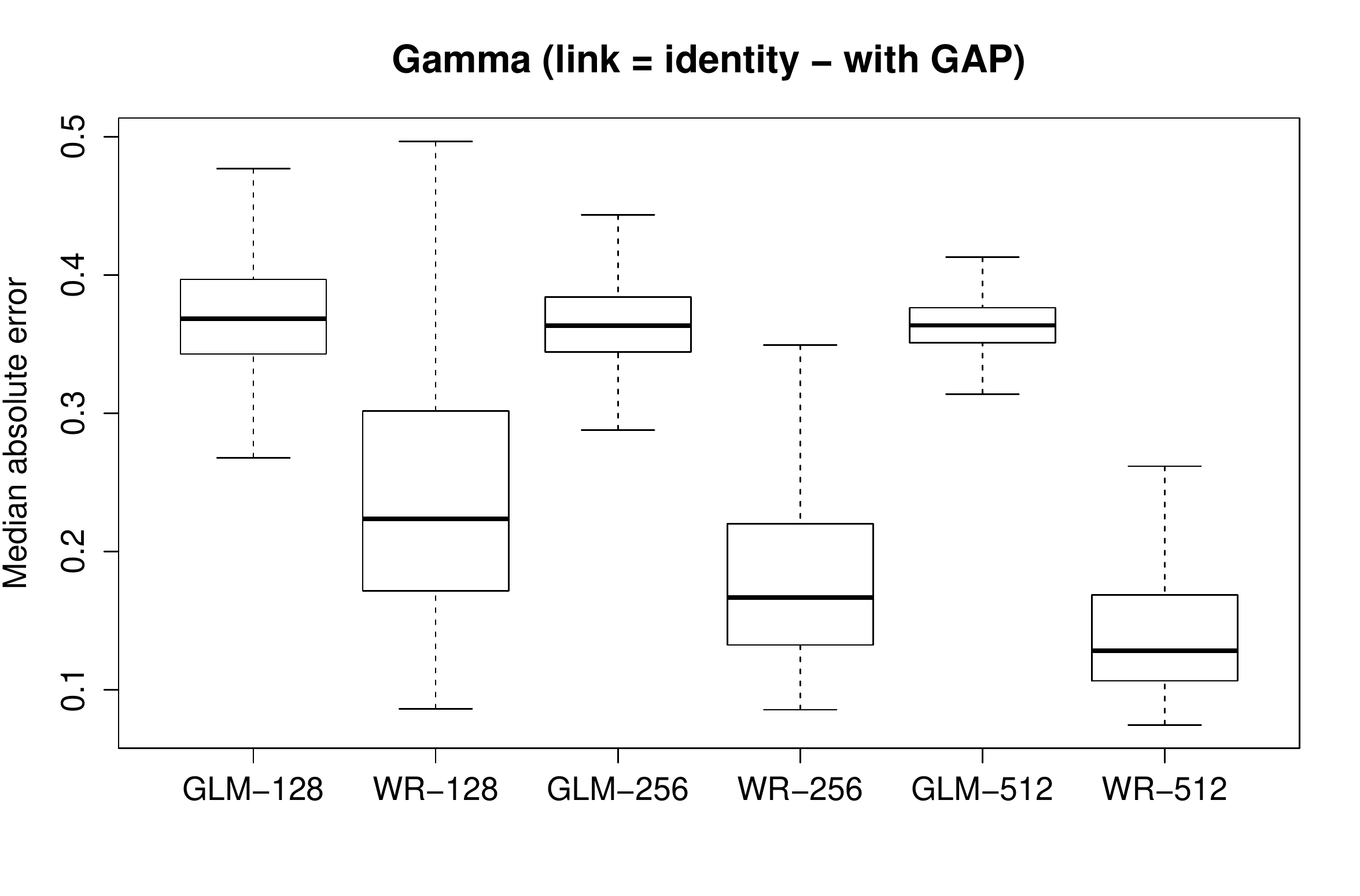}}
	\caption{Comparison between wavelet regression and GLM based on MAE.  Synthetic data sets with error gamma and link function identity link with gap. Dependence level weak (a), moderate (b) and strong (c).}
	\label{fig:Fig_Gama_identity_levels_MAE_GAP}
\end{figure}

Figure \ref{fig:Fig_Gama_identity_levels_MAE} illustrates the MAE obtained in the Monte Carlo simulation for the wavelet regression and the true GLM (fitted values), taking into account synthetic data sets with error gamma and link function identity. The first plot represents the box-plots for a weak dependence level between $Y$ and $\eta$. It is possible to verify a small difference for the MAE between the wavelet regression and GLM in all sample sizes. However, the second and third plots demonstrate that the fitted GLM outperforms the wavelet regression. Thus, when the dependence level is moderate or strong the GLM presented a better performance in comparison with the wavelet model. We also observe that as larger is the sample size as lower is the difference between the approaches. On the other hand, Figure \ref{fig:Fig_Gama_identity_levels_MAE_GAP} compares the performance of the wavelet regression and the GLM for data with the presence of a gap in the link function. Now, we conclude that the wavelet regression outperforms the GLM when the dependence level is moderate or strong.

From the results presented in Table \ref{tab:GLMoutperformWM} we verified that the GLM exhibited a better predictive performance in comparison with the wavelet model. The percentage of times that the GLM outperformed the wavelet regression is always higher than 60\%. 
\begin{table}
	\caption{Scenario 4 - Proportion of times that the GLM outperform the wavelet model according to the types of random component, link function, level of linkage, sample size and criterion (RMSE and MAE). Data sets without gap.}
	{\footnotesize
		\begin{center}
			\begin{tabular}{c|c|c|ccc|ccc}
				\hline \multirow{3}{*}{Random Comp.} & \multirow{3}{*}{Link} & \multirow{3}{*}{Level} &  \multicolumn{3}{c|}{$RMSE$} & \multicolumn{3}{c}{$MAE$} \\ \cline{4-9}
				&   &  & \multicolumn{3}{c|}{$n$} & \multicolumn{3}{c}{$n$} \\ 
				& 	 &  & 128  & 256  & 512   & 128 & 256  & 512   \\ \hline
				\multirow{9}{*}{Gaussian} & \multirow{3}{*}{Identity} & Weak & 0.942 & 0.974 & 0.976 & 0.728 & 0.733 & 0.763  \\
				&  & Moderate & 0.997 & 0.995 & 0.995 & 0.909 & 0.877 & 0.879  \\
				&  & Strong & 1.000 & 1.000 & 1.000 & 1.000 & 1.000 & 1.000  \\ \cline{2-9}
				&   \multirow{3}{*}{Inverse} & Weak & 0.938 & 0.964 & 0.978 & 0.698 & 0.680 & 0.694 \\
				&    & Moderate & 0.921 & 0.969 & 0.985 & 0.712 & 0.738 & 0.757 \\
				&   & Strong & 1.000 & 1.000 & 1.000 & 0.990 & 0.994 & 0.991 \\ \cline{2-9}
				&   \multirow{3}{*}{Log} & Weak & 0.979 & 0.987 & 0.967 & 0.859 & 0.831 & 0.806 \\
				&    & Moderate & 0.993 & 0.984 & 0.964 & 0.966 & 0.923 & 0.883 \\
				&    & Strong & 1.000 & 0.998 & 0.966 & 0.990 & 0.974 & 0.930 \\\hline 
				\multirow{9}{*}{Gama} &  \multirow{3}{*}{Identity}  & Weak & 0.948 & 0.973 & 0.968 & 0.746 & 0.761 & 0.750  \\
				&    & Moderate & 1.000 & 1.000 & 1.000 & 0.978 & 0.968 & 0.962  \\
				&   & Strong & 1.000 & 1.000 & 1.000 & 1.000 & 1.000 & 1.000  \\ \cline{2-9}
				&   \multirow{3}{*}{Inverse} & Weak & 0.944 & 0.975 & 0.978 & 0.747 & 0.761 & 0.751 \\
				&    & Moderate & 1.000 & 1.000 & 1.000 & 0.983 & 0.976 & 0.970 \\
				&    & Strong & 1.000 & 1.000 & 1.000 & 1.000 & 1.000 & 1.000 \\ \cline{2-9}
				&   \multirow{3}{*}{Log} & Weak & 0.786 & 0.876 & 0.926 & 0.636 & 0.603 & 0.709 \\
				&    & Moderate & 0.965 & 0.970 & 0.980 & 0.890 & 0.880 & 0.869 \\
				&    & Strong & 1.000 & 1.000 & 1.000 & 0.999 & 0.999 & 1.000 \\ \hline
				\multirow{12}{*}{Inverse Gaussian} & \multirow{3}{*}{Identity}  & Weak &  0.866 & 0.922 & 0.953 & 0.733 & 0.735 & 0.717  \\
				&  & Moderate &  0.979 & 0.976 & 0.987 & 0.935 & 0.897 & 0.897  \\
				&   & Strong &  0.999 & 0.999 & 0.998 & 0.994 & 0.990 & 0.990  \\ \cline{2-9}
				&  \multirow{3}{*}{Inverse} & Weak & 0.907 & 0.939 & 0.970 & 0.862 & 0.866 & 0.879 \\
				&   & Moderate & 0.985 & 0.981 & 0.989 & 0.954 & 0.929 & 0.928 \\
				&   & Strong & 1.000 & 0.999 & 1.000 & 0.996 & 0.994 & 0.988 \\ \cline{2-9}
				&   \multirow{3}{*}{Log} & Weak & 0.846 & 0.943 & 0.985 & 0.853 & 0.836 & 0.819 \\
				&    & Moderate & 0.974 & 0.974 & 0.997 & 0.978 & 0.978 & 0.975 \\
				&    & Strong & 0.998 & 0.998 & 1.000 & 1.000 & 1.000 & 0.999 \\ \cline{2-9}
				&   \multirow{3}{*}{$1/\mu^2$} & Weak & 0.936 & 0.962 & 0.965 & 0.779 & 0.784 & 0.808 \\
				&    & Moderate & 0.996 & 0.993 & 0.996 & 0.938 & 0.902 & 0.894 \\
				&    & Strong & 0.998 & 0.999 & 1.000 & 0.977 & 0.949 & 0.941 \\ \hline
			\end{tabular}
			\label{tab:GLMoutperformWM}
		\end{center}
	}
\end{table}
Table \ref{tab:GLMoutperformWMGAP} compares the predictive performances of the GLM and wavelet regression model due to the presence of a link gap, as illustrated in Figure \ref{fig:Fig_level_dependences_Gaussian_Log_GAP}. In this situation, the results demonstrated that the wavelet model outperforms the GLM when the dependence level is strong and when the sample size increases. For a weak level of dependence and small sample size the the GLM' and WR' predictive performances are quite similar.   
\begin{table}
	\caption{Scenario 4 - Proportion of times that the GLM outperform the wavelet model according to the types of random component, link function, level of linkage, sample size and criterion (RMSE and MAE). Data sets with gap.}
	{\footnotesize
		\begin{center}
			\begin{tabular}{c|c|c|ccc|ccc}
				\hline \multirow{3}{*}{Random Comp.} & \multirow{3}{*}{Link} & \multirow{3}{*}{Level} &  \multicolumn{3}{c|}{$RMSE$} & \multicolumn{3}{c}{$MAE$} \\ \cline{4-9}
				&   &  & \multicolumn{3}{c|}{$n$} & \multicolumn{3}{c}{$n$} \\ 
				& 	 &  & 128  & 256  & 512   & 128 & 256  & 512   \\ \hline
				\multirow{9}{*}{Gaussian} & \multirow{3}{*}{Identity} & Weak & 0.614 & 0.407 & 0.156 & 0.513 & 0.346 & 0.163  \\
				&  & Moderate & 0.535 & 0297 & 0.119 & 0.285 & 0.079 & 0.004  \\
				&  & Strong & 0.499 & 0.272 & 0.109 & 0.166 & 0.054 & 0.001  \\ \cline{2-9}
				&   \multirow{3}{*}{Inverse} & Weak & 0.835 & 0.776 & 0.667 & 0.607 & 0.556 & 0.458 \\
				&    & Moderate & 0.508 & 0.316 & 0.104 & 0.438 & 0.343 & 0.201 \\
				&   & Strong & 0.222 & 0.065 & 0.009 & 0.108 & 0.033 & 0.003 \\ \cline{2-9}
				&   \multirow{3}{*}{Log} & Weak & 0.540 & 0.278 & 0.090 & 0.214 & 0.061 & 0.005 \\
				&    & Moderate & 0.519 & 0.263 & 0.084 & 0.101 & 0.022 & 0.000 \\
				&    & Strong & 0.514 & 0.264 & 0.085 & 0.070 & 0.018 & 0.000 \\\hline 
				\multirow{9}{*}{Gama} &  \multirow{3}{*}{Identity}  & Weak & 0.591 & 0.408 & 0.162 & 0.509 & 0.359 & 0.167  \\
				&    & Moderate & 0.520 & 0.277 & 0.112 & 0.184 & 0.037 & 0.003  \\
				&   & Strong & 0.455 & 0.241 & 0.119 & 0.137 & 0.060 & 0.010  \\ \cline{2-9}
				&   \multirow{3}{*}{Inverse} & Weak & 0.307 & 0.127 & 0.014 & 0.311 & 0.233 & 0.071 \\
				&    & Moderate & 0.185 & 0.040 & 0.006 & 0.160 & 0.033 & 0.004 \\
				&    & Strong & 0.168 & 0.031 & 0.004 & 0.133 & 0.041 & 0.005 \\ \cline{2-9}
				&   \multirow{3}{*}{Log} & Weak & 0.393 & 0.259 & 0.101 & 0.406 & 0.274 & 0.124 \\
				&    & Moderate & 0.365 & 0.138 & 0.033 & 0.179 & 0.045 & 0.007 \\
				&    & Strong & 0.365 & 0.134 & 0.034 & 0.074 & 0.016 & 0.004 \\ \hline
				\multirow{12}{*}{Inverse Gaussian} & \multirow{3}{*}{Identity}  & Weak &  0.596 & 0.499 & 0.413 & 0.313 & 0.161 & 0.046  \\
				&  & Moderate &  0.394 & 0.178 & 0.070 & 0.084 & 0.018 & 0.000  \\
				&   & Strong &  0.348 & 0.146 & 0.046 & 0.035 & 0.006 & 0.000  \\ \cline{2-9}
				&  \multirow{3}{*}{Inverse} & Weak & 0.551 & 0.483 & 0.291 & 0.462 & 0.418 & 0.229 \\
				&   & Moderate & 0.254 & 0.083 & 0.012 & 0.314 & 0.080 & 0.015 \\
				&   & Strong & 0.176 & 0.054 & 0.007 & 0.197 & 0.063 & 0.007 \\ \cline{2-9}
				&   \multirow{3}{*}{Log} & Weak & 0.791 & 0.878 & 0.950 & 0.467 & 0.389 & 0.192 \\
				&    & Moderate & 0.502 & 0.378 & 0.246 & 0.124 & 0.029 & 0.003 \\
				&    & Strong & 0.249 & 0.076 & 0.011 & 0.120 & 0.003 & 0.002 \\ \cline{2-9}
				&   \multirow{3}{*}{$1/\mu^2$} & Weak & 0.500 & 0.321 & 0.091 & 0.414 & 0.343 & 0.190 \\
				&    & Moderate & 0.250 & 0.069 & 0.010 & 0.264 & 0.076 & 0.006 \\
				&    & Strong & 0.200 & 0.057 & 0.009 & 0.200 & 0.054 & 0.005 \\ \hline
			\end{tabular}
			\label{tab:GLMoutperformWMGAP}
		\end{center}
	}
\end{table}

Finally, based on the results of the \textcolor{black}{Tables \ref{tab:trueclas}}, \ref{tab:GLMoutperformWM}  and \ref{tab:GLMoutperformWMGAP}, we concluded that the WP is an important tool to identify the best nonlinear regression structure or the best link function for a GLM. However, one time chosen the more appropriate parametric regression structure by the WP , the chosen parametric model will provide the best predictive values for the response variable $Y$. The exception occurs due the presence of some atypical behavior in the data as, for example, the presence of gaps.


\section{\textcolor{black}{Application to a real data set}}

\textcolor{black}{This section brings two applications to real data sets. The aim is to evaluate the proposed WP in real problems. The first application uses the WP to find the more adequate link function for a GLM. The second example use the WP to identify the more appropriate nonlinear relationship from a range of 26 candidate's models.}

\subsection {\textcolor{black}{Semiconductor manufacturing process data set}} \label{ME}

\begin{figure}
	\centering
	\includegraphics[width=0.4\linewidth]{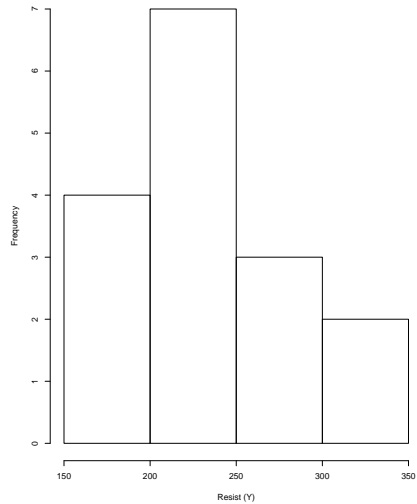}
	\caption{Empirical distribution of the variable resistivity.}
	\label{fig:hist_resist}
\end{figure}

The data set consists of the a semiconductor manufacturing process. 
It is believe that four factors influence the resistivity ($Y$) of the wafer, so a full factorial experiment with two levels for each factor is designed and employed. Previous analysis conclude that a Box-Cox yields a log transformation normal model for $Y$. However,  \cite{Faraway2006} concludes that a GLM gamma model with ``log'' link function is better than the transformed linear normal model based on AIC criterion.  Figure \ref{fig:hist_resist} illustrates the presence of right asymmetry in the empirical distribution of the response variable $Y$, which corroborates the assumption of a Gamma distribution in the random component of the GLM. 

\begin{figure}
	\centering
	\includegraphics[width=0.6\linewidth]{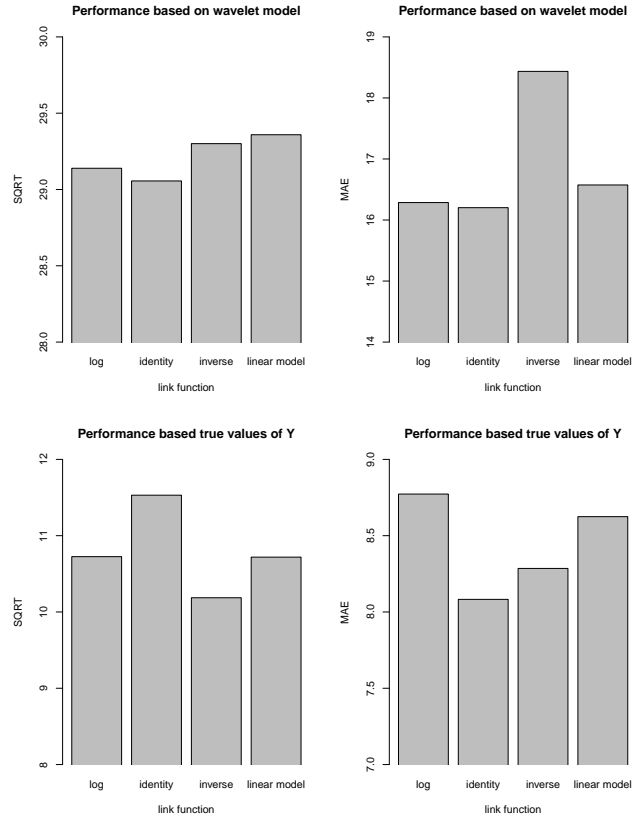}
	\caption{Choice of the link function choice based on WP, according to the criterion.}
	\label{fig:LF_wavelet}
\end{figure}

However, according with the proposed WP we suggests that the link function ``identity'' presents a better fit for this data set, according with the measures RMSE and MAE (Figure \ref{fig:LF_wavelet}). Notice that the GLM gamma with link function ``log'' presented an intermediate performance while the linear model demonstrated the worst fit. The GLM gamma with``inverse'' link function presented the worst performance for the MAE criterion. The two bar plots on the upper half of the Figure \ref{fig:LF_wavelet} illustrate these results.

We also repeat the previous analysis replacing the predicted values of the wavelet by the true values of $Y$. The GLM gamma with ``log'' link function and the linear model presented similar performance based on the RSME criterion. However, the GLM gamma model with ``identity'' link function presented the best fit based on the MAE criteria. The two bar plots on the bottom half of the Figure \ref{fig:LF_wavelet} illustrate these results.

\begin{figure}
	\centering
	\includegraphics[width=0.4\linewidth]{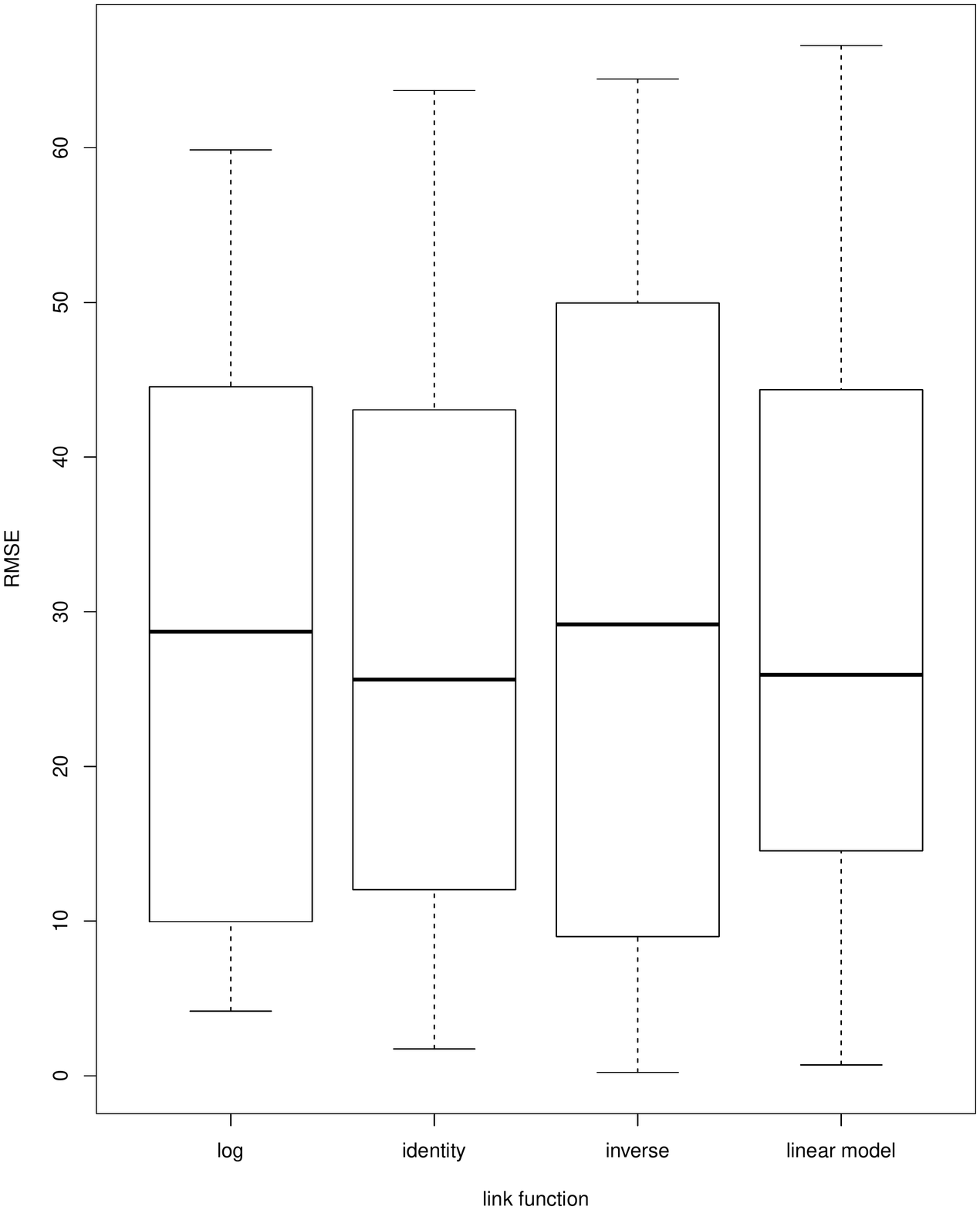}
	\caption{Predictive performance between the models based on leave-one-out scheme.}
	\label{fig:pred_performance}
\end{figure}

Moreover, we also evaluate the predictive performance of the four models based on a leave-one-out scheme. Figure \ref{fig:pred_performance} depicts that all models presents a very similar behavior.  However, the GLM model with ``identity'' link function presents the lower median error and a low variability, according to the box-plots.  We remember that the MAE criterion presented the best accuracy rate to identify the more appropriate link function for a GLM in the simulation section. Furthermore, the response and explanatory variables are not transformed in the GLM gamma model with ``identity'' link function, producing a very easy interpretation for the parameter estimates. Based on these facts, we believe that the GLM with ``identity'' link function is the most appropriate model for this data set.

\subsection{Rabbits in Australia data set}

This section brings an application to a real data set, originally, presented by \cite{DM1961} and later studied by \cite{Ratkowsky1983}. The authors suggest the use of a nonlinear regression model to study the relation between variables dry weight of eye lens ($Y$) and age of rabbit ($X$). The study consider 71 European rabbit ({\it Oryctolagus cuniculus} in Australia. The true nonlinear regression model is denoted by the expression $f_{26}$, available in the supplementary material.

According with the results presented in Table \ref{tcoelhos}, the WP suggests the model $f_{26}$ as best one among the 26 candidate's models. This mean that the fitted values of the true parametric model is the more close of the fitted values of the wavelet model. Notice that, according to both criteria, the model $f_{26}$ will be suggested as the true model by the WP.
\begin{table}
	\centering
	\caption{Choice of the nonlinear equation based on RMSE and MAE criteria.} \label{tcoelhos}
	\begin{tabular}{cccccc}
		\hline
		Model & RMSE & EAM & Model & RMSE & EAM \\ 
		\hline
		$f_1$ & 0.48 & 0.44 & $f_{14}$ & 4.03 & 4.17 \\ 
		$f_2$ & 0.48 & 0.44 & $f_{15}$ & 4.89 & 5.17\\ 
		$f_3$ & 1.16 & 0.76 & $f_{16}$ & 7.40 & 2.74 \\ 
		$f_4$ & 5.01 & 5.09 & $f_{17}$ & 0.39 & 0.37\\ 
		$f_5$ & 5.03 & 5.17 & $f_{18}$ & 3.04 & 3.25 \\ 
		$f_6$ & 2.84 & 2.62 & $f_{19}$ & 5.03 & 5.17 \\ 
		$f_7$ & 2.81 & 2.66 & $f_{20}$ & 0.76 & 0.57 \\ 
		$f_8$ & 4.03 & 4.17 & $f_{21}$ & 5.00 & 5.16 \\ 
		$f_9$ & 4.03 & 4.17 & $f_{22}$ & 5.03 & 5.17 \\ 
		$f_{10}$ & 4.03 & 4.17 & $f_{23}$ & 5.03 & 5.17 \\ 
		$f_{11}$ & 4.03 & 4.17 & $f_{24}$ & 0.32 & 0.22 \\ 
		$f_{12}$ & 4.03 & 4.17 & $f_{25}$ & 5.03 & 5.17 \\ 
		$f_{13}$ & 4.03 & 4.17 & \textbf{\textcolor{black}{$f_{26}$}} & \textbf{\textcolor{black}{0.25}} & \textbf{\textcolor{black}{0.12}}\\ 
		\hline
	\end{tabular}
\end{table}

\begin{figure}
	\centering
	\caption{Empirical relationship between $X$ e $Y$ and predicted values of the parametric model $f_{26}$ and wavelet regression.}\label{fcoelhos}
	\includegraphics[width=12cm]{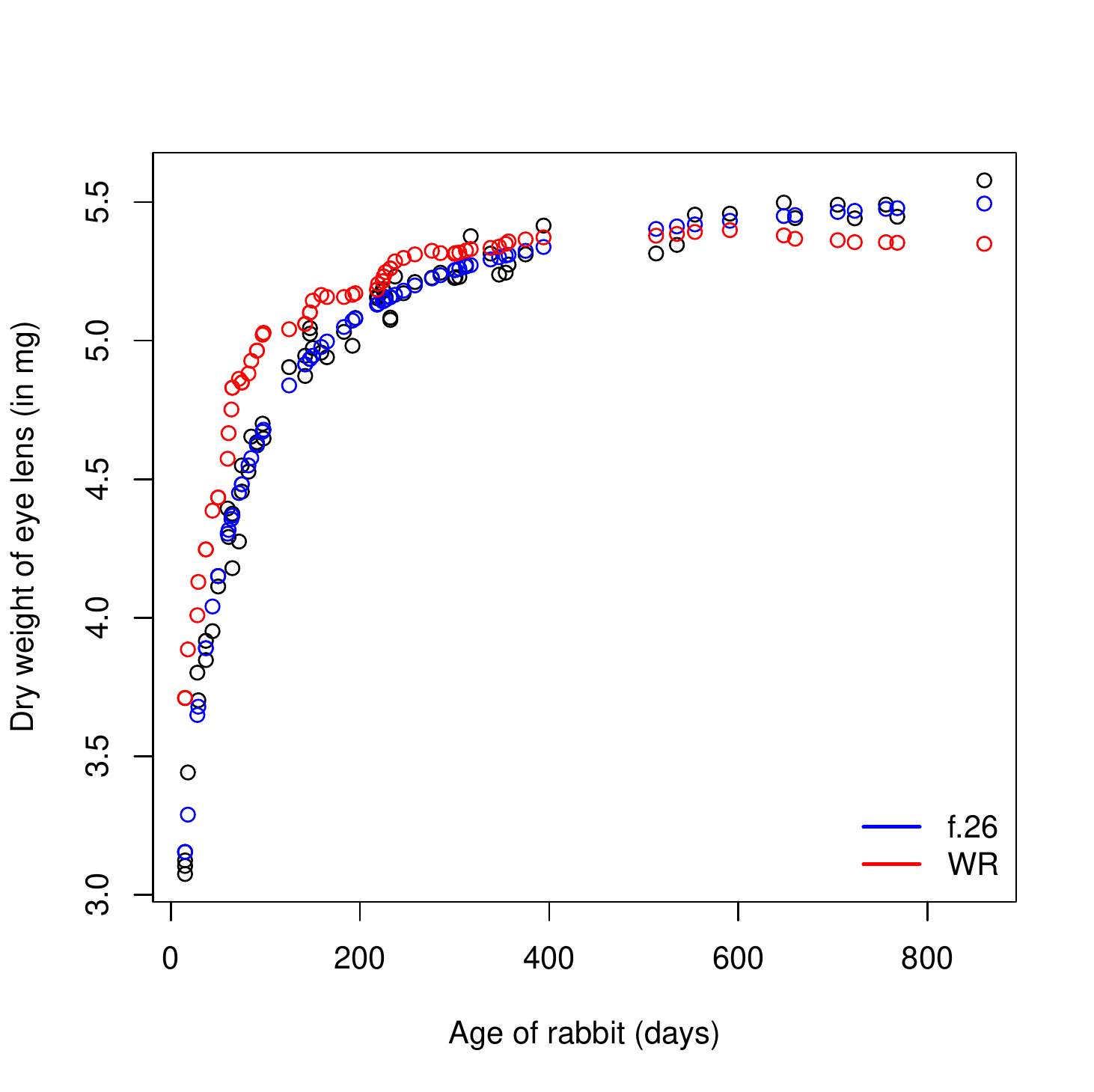}
\end{figure}

Finally, Figure \ref{fcoelhos} illustrates the nonlinear relationship between the variables $Y$ and $X$ (black points), the fitted values of the wavelet model (red points) and the fitted values according with the chosen nonlinear function $f_{26}$ (blue points). Notice that the blue points are very close to the true values which suggests that nonlinear model $f_{26}$ represents a good model for this data set. 

\section{\textcolor{black}{Concluding remarks}}

In this paper we proposed the use of a wavelet procedure (WP) to identify the best parametric model among a list of $k$ possible candidate's models. We considered the case of the choice of the best nonlinear equation and the case of the best link function in the GLM context. Initially, the procedure fits a non-parametric wavelet regression model to the data. Thus, we perform a comparison between the wavelet model and each one of the $k$ parametric models, considering a predefined performance error measure (like RMSE or MAE) taking into account the fitted values of both models. The procedure identifies the parametric model more close to the wavelet regression. Thus, this model is considered as the ``best'' parametric model for the data.

A experimental study based on Monte Carlo framework was proposed to evaluate the accuracy of the WP to identify the true parametric model. The results were obtained in terms of true classification rate of the WP, taking into account 4 different scenarios, 3 different sample sizes, 3 different dependence level between $Y$ and $X$, among others aspects, in a total of 138 different configurations. We considered a total of $25$ candidate's models.

The WP presented a high true classification rate to detect the true nonlinear parametric model in scenarios 1 and 2, even when the competitor model presents a very similar behavior in relation to the true model. We also verified that the WP detects the more appropriate link function for a GLM model (scenario 3). Moreover, the wavelet regression model considered few non null coefficients, suggesting that the non-parametric model does not overfitting the data. Although of the good accuracy of the WP to detect the true parametric form of a regression model, the fitted values of the parametric model presented a lower residual (for the true values) in comparison with the fitted values of the wavelet regression, when the dependence level between $Y$ and $X$ is moderate or strong. On the other hand, for data sets with the presence of gap the wavelet model outperformed the parametric model in terms of fitted values. 

%

The experimental results suggest that the WP is an important tool to identify the best nonlinear function or the best link function. However, one time chosen the more appropriate parametric form by the WP, the chosen parametric model provided the best predictive values for the response variable $Y$. The exception occurred due the presence of some atypical behavior in the data as, for example, the presence of gaps.

The applications to a real data sets corroborate the results obtained in the simulation section and demonstrated the usefulness of the WP to choose a appropriate parametric form for a regression model in terms of nonlinear regression and generalized linear model. 

%

\section*{Acknowledgements}
The second author  acknowledges FAPESP (Funda\c{c}\~ao de Amparo \`a Pesquisa do Estado de S\~ao Paulo) grant number 2013/00506-1 and CNPq (Conselho Nacional de Desenvolvimento Cient\'{\i}fico e Tecnol\'ogico) grant number 308439/2014-7.

%

\bibliographystyle{chicago}

\end{document}